\documentclass[useAMS,usenatbib]{mn2e}

\usepackage[english]{babel}
\usepackage{amssymb}
\usepackage{enumerate}
\usepackage{epsfig,amsmath}
\usepackage{wrapfig}
\usepackage{sidecap}
\usepackage{graphics}
\usepackage{rotating}
\usepackage{boldtensors}

\title[PCA for variable stars in SDSS Stripe 82]{Search for high-amplitude $\delta$ Scuti  and RR Lyrae stars in Sloan Digital Sky Survey Stripe 82 using principal component analysis}
\author[M. S\"uveges et al.]{M. S\"uveges$^{1,2}$\thanks{E-mail:
Maria.Suveges@unige.ch},  B. Sesar$^{3}$, M. V\'aradi$^{2}$, N. Mowlavi$^{1,2}$, A. C. Becker$^{4}$, \v{Z}. Ivezi\'c$^{4}$, \newauthor M. Beck$^{1,2}$, K. Nienartowicz$^{1,2}$, L. Rimoldini$^{1,2}$, P. Dubath$^{1,2}$, P. Bartholdi$^{2}$,  \newauthor and L. Eyer$^{2}$ \\
$^{1}$ISDC Data Centre for Astrophysics, University of Geneva, Chemin d'Ecogia 16, CH-1290 Versoix, Switzerland\\
$^{2}$Department of Astronomy, University of Geneva, Chemin des Maillettes 51, CH-1290 Sauverny, Switzerland\\
$^{3}$Division of Physics, Mathematics and Astronomy, California Institute of Technology, Pasadena, CA 91125, USA \\
$^{4}$University of Washington, Department of Astronomy, P.O. Box 351580, Seattle, WA 98195-1580, USA
}
\begin{document}

\date{}

\pagerange{\pageref{firstpage}--\pageref{lastpage}} \pubyear{2012}

\maketitle

\label{firstpage}

\begin{abstract}
We propose a robust principal component analysis (PCA) framework for the exploitation of multi-band photometric measurements in large surveys. Period search results are improved using the time series of the first principal component due to its optimized signal-to-noise ratio.The presence of correlated excess variations in the multivariate time series enables the detection of weaker variability.  Furthermore, the direction of  the largest variance differs for certain types of variable stars. This can be used as an efficient attribute for classification. The application of the method to a subsample of Sloan Digital Sky Survey Stripe 82 data yielded 132 high-amplitude $\delta$ Scuti variables. We found also 129 new RR Lyrae variables, complementary to the catalogue of \citet{sesaretal10}, extending the halo area mapped by Stripe 82 RR Lyrae stars towards the Galactic bulge. The sample comprises also 25 multiperiodic or Blazhko RR Lyrae stars. 
\end{abstract}


\section{Introduction}

During the last decade, wide-area and multi-epoch surveys have started to play a major role in astronomical research. Developments in astronomical instrumentation and in space observation techniques, together with the rapidly growing data storage facilities and the broadly available software for combining these data
provide an enormous wealth of information. The traditional manual procedures must be replaced by quick automated methods for pre-processing, selection and analysis. 

Analysis of variable stars has benefited greatly from data collected by large-scale surveys such as ASAS \citep{pojmanski02, pojmanski03}, OGLE \citep{udalskietal96}, MACHO \citep{alcocketal97}, or EROS \citep{aubourgetal93,spanoetal11}. Studies of different, sometimes rare classes of objects become possible with unprecedentedly large sets of objects. These studies require specific pre-processing: a preliminary classification of the observed objects, in order to enable an efficient extraction of homogeneous samples. The first step in this procedure is variability detection, providing a set of candidate variable stars. The next is characterization of the observed time series by the calculation of some numerical summaries of the observed time series. These are usually statistical moments (e.g. mean magnitude, skewness, kurtosis), derived quantities from period search (e.g. amplitudes and relative phases of harmonic components), and some astrophysical parameters such as colours. These parameters, called generally attributes in the context of classification, are then used to estimate the types of the objects. The volume of data requires fast automated data mining techniques like, for instance, those proposed by \citet{eyerblake02,eyerblake05}. Most recently, Random Forest \citep{breiman01,dubathetal11, richardsetal11,rimoldinietal12}, multistage Bayesian networks \citep{debosscheretal07, sarroetal09} and gradient boosting methods \citep{richardsetal11} were tested for this purpose with promising results. 

Unfortunately, automatically distributed class labels can be less reliable than the results of a careful human inspection. A more efficient use of the information contained in the data can improve on this. The goal of this study is to consider a new way of including colour information into automated classification procedures. Although variable stars with different origin of variability and different physical parameters show different amplitudes and light curve patterns depending on wavelength during their cycle, this variation is not easy to summarize in a concise numerical form, as it is a function of phase. We attempt to find summaries of these colour variations, and to use it as complementary information to the usual attributes in automated classification of multi-band survey data.

Principal component analysis (PCA, \citealt{jolliffe,hastieetal}) gives precisely such summaries. It considers the vector of observations in $M$ bands at one time as a point in an $M$-dimensional space, and looks for a decomposition of the space of observations into directions with maximal variance. The first direction is called the first principal component or  PC1. The projection of any point to the PC1 direction is a linear combination of the simultaneous measurements with constant coefficients, which is called PC1 score. It provides several potential advantages. It improves signal-to-noise ratio when searching for periods on the time series of the projections on the direction of PC1. It yields a variability criterion based on the presence of correlated variations across filters. Also, the direction of PC1 is characteristic to the origin of variability and to the physical properties of the star, and is useful to separate eclipsing binaries from some pulsating variables with symmetric light curves such as RR Lyrae type c (hereafter RRc). We demonstrate these advantages on 5-band time series from a flux-limited sample of Stripe 82 objects of the Sloan Digital Sky Survey (SDSS). 

However, applying PCA on astronomical time series is not straightforward. Usually, the data in different filters have different error levels, depending on the measured magnitudes too. Outliers and non-normality can also strongly influence PCA, which is built entirely on the assumption of normality. Since usually not all type of errors can be anticipated and accounted for in advance, the robustness of the methods is very desirable: a good automated method should be able to function acceptably well even in the presence of contaminations from various unknown sources. We propose a combination of variance stabilizing transformation with robust principal component analysis to minimize the impact of these issues.  

After constructing the methodology, we use it to select  samples of candidate RR Lyrae and high-amplitude $\delta$ Scuti (HADS) from SDSS Stripe 82. RR Lyrae are interesting as they are Population II halo structure tracers, and obey well-determined period-luminosity relations (e.g. \citealt{smith}). The HADS stars are also pulsating variable stars with spectral type from late A to early F, occurring in the instability strip on and just above the main sequence below the RR Lyrae (e.g. \citealt{breger80,clementetal01,pigulskietal06}). Their pulsation is in majority radial-mode, though for some, there are indications of non-radial modes \citep{mazuretal03,poretti03}. They also satisfy period-luminosity relations depending on metallicity and oscillation mode \citep{nemecetal94,mcnamara95,mcnamara97,mcnamara00}, but Population I ($\delta$ Scuti) and II (SX Phoenicis) objects are mixed in the group, and they are affected also by mode identification problems. In Stripe 82 data, the confirmed RR Lyrae stars published by \citet{sesaretal10} will be used as a performance test and a training set for the PCA-based methodology. For HADS stars, we build a training set to obtain a clean sample in Stripe 82. Finally, we present a list of 129 new RR Lyrae and 132 HADS candidate variables.

In Section \ref{sec-SDSS}, we briefly present SDSS Stripe 82 \citep{yorketal00}. Section \ref{sec_theory} summarizes the statistical background: variance stabilizing transformation and principal component analysis with its robust version, the tested period search variants, and the Random Forest classifier. Section \ref{sec_appl} describes the results obtained by PCA in variability detection, period search, characterisation and classification. A short summary and a discussion follows in Section \ref{sec_disc}.

\section{SDSS Stripe 82} \label{sec-SDSS}

The Sloan Digital Sky Survey Data Release 7 \citep{abazajianetal09} provides five-band ($u$, $g$, $r$, $i$ and $z$) photometry of more than 11000 deg$^2$ of the sky. The photometric errors are around 0.02 mag for $g < 16$ mag, and around $ 0.04$ mag for $g\sim$  21 mag (see Fig. 2 of \citealt{sesaretal10}). Equivalent values for the noisiest $u$ band are 0.02 mag at the bright end and 0.05 mag around 20 mag. The 95\% completeness limits of the images are $u = 22.0$, $g = 22.2$, $r = 22.2$, $i = 21.3$, $z = 20.5$ \citep{abazajianetal04}. One of the southern regions, Stripe 82, was observed repeatedly during the first phase of SDSS (SDSS-I) and the following SDSS-II Supernova Survey \citep{bramichetal08, friemanetal08}. The measurements in the five bands were taken quasi-simultaneously, with around 1.2 minute time difference between band records. 

Our data set is a flux-limited subset of \citet{sesaretal07}, separated into around 68,000 variable and 200,000 non-variable objects with median $g$ magnitude brighter than 20.5 mag. The variable flag used there was based on two criteria: (1) for the root mean square scatter in $g$ and $r$, $\sigma_r > 0.05$ and $\sigma_g > 0.05$; (2) for the reduced chi-square in $r$ and  $g$,  $\xi_r \geq 3$ and   $\xi_g \geq 3$, respactively. This selection will be referred as `flagged as variable' or `flagged non-variable' hereafter, and will serve in comparisons with the PCA-based detection method. Details about the construction and testing of the catalogue are described in  \citet{ivezicetal07}. \citet{sesaretal10} published 483 confirmed RR Lyrae variables in Stripe 82; our subset of data contains 450 of these. This makes Stripe 82 data particularly adapted to test our methods, as we can check their performance by direct comparison to previous analyses of the same data.

\section{Statistical tools}\label{sec_theory}

\subsection{Principal Component Analysis}\label{subsec_PCA}

Principal component analysis is a statistical tool developed for finding orthogonal directions of maximal variance in high-dimen\-si\-onal data sets. It is assumed that directions of large variance are of particular interest: in signal processing, they may contain the bulk of the information transmitted by a signal; in image analysis, they may concisely summarise the most characteristic pattern forms; or in classification, they might coincide with the directions that best separate some distinct groups of objects. In order to find them, we consider the point cloud of multi-filter observations plotted against each other, regardless of times. The goal is to find an orthogonal coordinate system adapted to this point cloud: the first axis is fixed so that it points to the direction of maximal variance. Next, the points are decomposed into projections to this first axis and to its orthogonal subspace. Then the second axis is chosen to point into the direction of largest variance in the orthogonal subspace. These steps are repeated until an orthogonal basis spanning the original space is found. Mathematically, we seek for a successive decomposition of the space of the observables into orthogonal directions to which the projections of the points have the highest residual variance. 

Suppose that we observed $M$ sequences 
$X_{11},X_{21}, \ldots, X_{N1};$ $ \ldots; \quad X_{1M}, X_{2M}, \ldots, X_{NM}$ at $N$ times, with the values $X_{i1},X_{i2}, \ldots, X_{iM}$ observed simultaneously. This corresponds to a sequence of observations made in an $M$-dimensional space, where a joint observation $X_{i1},X_{i2}, \ldots, X_{iM}$ is represented by a point in the $M$-dimensional state space. If we define the matrix  $\mathbf{X}$ so that its columns are the $M$ sequences, and the rows correspond to the different $M$-dimensional observations, then we can write the sample covariance matrix as $\mathbf{X}^T\mathbf{X} / N$, with the superscript $T$ denoting transposition. Finding orthogonal directions of maximal variance turns out to be equivalent to finding the eigendecomposition 
\begin{eqnarray*} 
 \mathbf{X}^T\mathbf{X} = \mathbf{V} \mathbf{D}^2 \mathbf{V}^T.
\end{eqnarray*}
Here the columns of $\mathbf{V}$ are the eigenvectors, or in other terms, the principal directions of $\mathbf{X}$, and $\mathbf{D}^2$ is a diagonal matrix with non-negative values $d_1^2, d_2^2, \ldots, d_M^2$, representing the variances of the projections of the points on the principal directions. By convention, the order of the directions is such that $d_1^2 > d_2^2 > \cdots > d_M^2$, and therefore the first principal component corresponds to the direction of the maximal variation. The vector of projections of the points on the first principal direction can be written as the linear combination $\mathbf{z}_1 = \mathbf{X}\mathbf{v}_1$, where $\mathbf{v}_1$ is the first column of the matrix $\mathbf{V}$.

\begin{figure}
\begin{center}
\includegraphics[scale=.83]{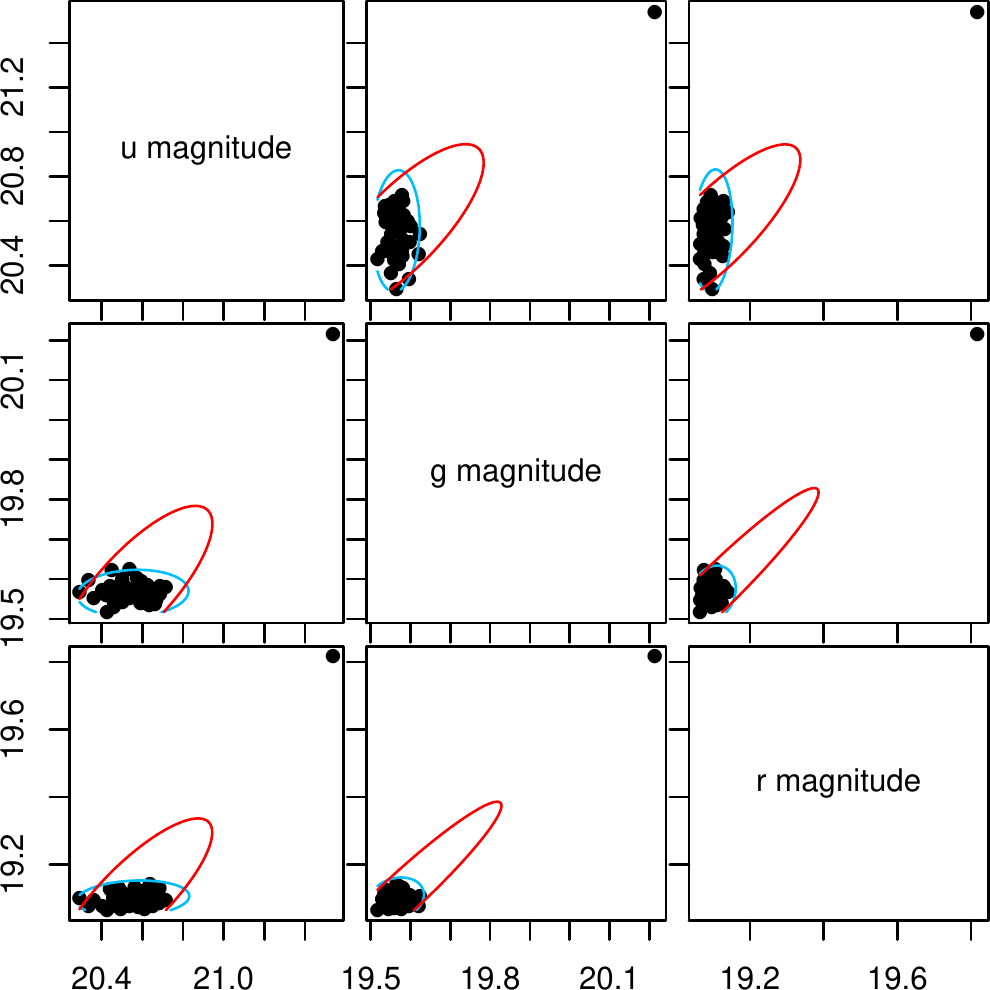}
\caption{ $u$-, $g$- and $r$-magnitudes of a randomly chosen star flagged as variable from SDSS Stripe 82. The red ellipsoids are levels of equal probability density fitted by non-robust maximum likelihood, the blue ones are the same levels from a robust fit by the minimum covariance determinant method.}
\label{fig:pairplot}       
\end{center}
\end{figure}

How to apply this for variable star analysis? For $M$ time series of a star consisting of $N$ simultaneously taken points, we consider the $M$-dimensional point cloud of the observations, as shown in Fig. \ref{fig:pairplot}.  We intend to apply principal component analysis in order to find the direction where variability is maximal, so that we can find more easily the period of the variable star. However, the application immediately hits an obstacle visible in Fig. \ref{fig:pairplot}, the different average noise in the different bands. Even if a star shows coherent deterministic variations across the bands, this can be easily masked by the noise in one of the bands (in SDSS, the noisiest bands are generally the $u$ and the $z$ band). PCA will pick the direction of the noisiest band as shown by the blue ellipsoids in the upper middle and right panels of  Fig. \ref{fig:pairplot}, and not the direction of coherent variations. The remedy is to use the estimated errors to scale the observations so that we have unit variance in every band. In the case of a non-variable star and near-independent errors, this implies that the point cloud appears as a sphere. For a variable star, the noise is added to a deterministically varying light curve, which causes the centre of the sphere to move in the $M$-dimensional space, and instead of a ball, we observe an elongated ellipsoid-like shape. The first principal direction is the longest axis of this shape, providing the largest variability amplitude that can be obtained by a linear transformation. 

Assume that the distribution of the observations is $X_{im} \sim \mathcal{N}(\mu_m, \sigma_{im}^2)$, that is,  the star is constant with mean magnitude $\mu_m$ in band $m$, the standard deviation of the measurement $X_{im}$ around the mean is $ \sigma_{im}$, and the error distribution is normal. We have several options to obtain unit variance of the noise. 
\begin{enumerate}[(i)]
\setlength{\itemsep}{-0.05cm}
\item We can centre and scale the measurements by an estimate of the centre and the pointwise error estimates (called here local scaling):
$$Y_{im} = \frac{X_{im} - \hat \mu_m}{\sigma_{im}} \quad \quad \mathrm{for \, all } \quad m = 1, \ldots, M,$$
where $ \hat \mu_m$ can be either the mean or the median of $X_{im}$.  
\item We can centre and scale the measurements by an estimate of the centre and the sqare root of the average variance of the time series:
$$Y_{im} = \frac{X_{im} - \hat \mu_m}{\hat \sigma_m} \quad \quad \mathrm{for \, all } \quad m = 1, \ldots, M,$$
where $ \hat \mu_m$ and $\hat \sigma_m^2$ can be either the mean (non-robust scaling) or the median (robust scaling) of $X_{im}$ and of $\sigma_{im}^2$, respectively. In the case of correlated errors that can be assumed normal, an analogous matrix transformation can be based on the covariance matrix.  In this paper, we assume that the correlation between the errors is weak compared to the correlation between bands for an RR Lyrae or HADS star \citep{scrantonetal05}.
\item Considering that the errors depend on true magnitudes, and this effect can be strong for some classes of variable stars or very faint objects, we tested also  the variance stabilizing transformation \citep{everitt02}. Suppose we have a random variable $X$ from a distribution with mean $\mu$ and variance $\sigma^2$ for which $\sigma^2 = g(\mu)$. Define the function $f$ as the solution to $f'(x) = [g(x)]^{-1/2}$. Then the transformation $Y = f(X)$ leads to a variable with unit variance: $\mathrm{Var}(Y) = 1$. In our case, we supposed a functional form $ \sigma^2  = g(x) = a \exp( b x)$ between the variances $\sigma_{im}^2$ and centred observed magnitudes $x_{im}$  with different coefficients for each band. This corresponds to a rough approximation with Poisson noise, which leads to an easily tractable closed-form transformation. Fitting this function bandwise to observed magnitudes and squared standard errors of a sample of variable and non-variable stars resulted in the bandwise estimates $\hat a_m$ and $\hat b_m$ and a transformation $Y_{im} = 2 \hat  a_m^{-1/2} \hat b_m^{-1} \exp(-\hat b_m x_{im} /2)$ for band $m$. Then the point cloud of $Y_{im}$  is centred by removing either the mean or the median from it. 
\end{enumerate}

\begin{figure*}
\begin{center}
\includegraphics[scale=.78]{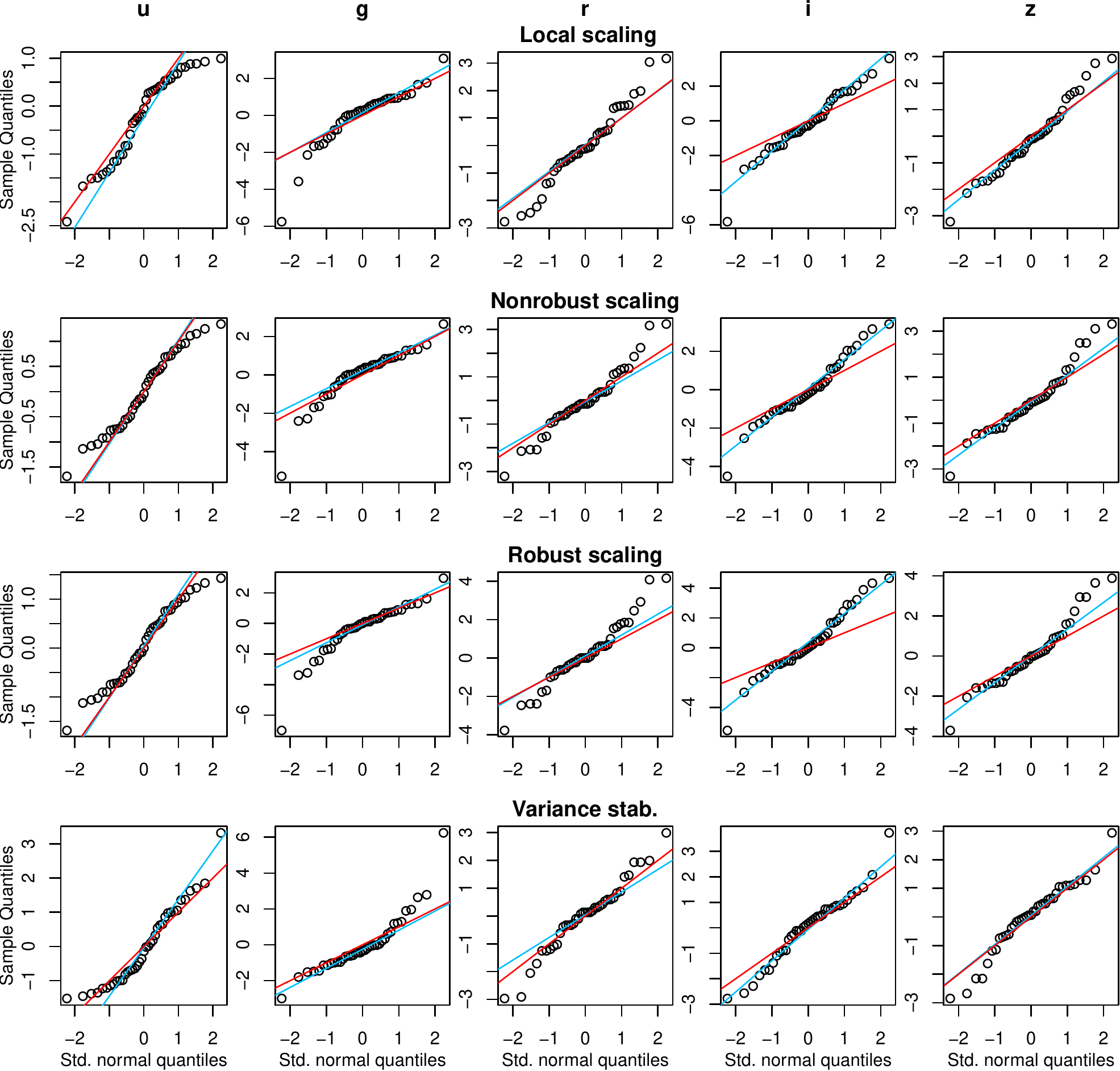}
\caption{Quantile-quantile plots of a non-variable star from Stripe 82. The columns are the five bands $u$,$g$,$r$,$i$,$z$. The rows correspond to different scaling methods: first row from top, local scaling; second row, non-robust scaling; third row, robust scaling; fourth row, variance stabilizing transformation. The red line corresponds to a standard normal distribution, the blue line to a robustly fitted best-fit normal distribution to the data.}
\label{fig:qq}       
\end{center}
\end{figure*}

The assumption of normality can be checked by quantile-quantile (Q-Q) plots \citep{chambersetal83, cleveland94}. After scaling and centring, the ordered sample of transformed magnitudes of a star is plotted against the corresponding theoretical quantiles of the standard normal distribution. If the sample follows indeed the standard normal distribution, then the points should be aligned along a straight line with intersect 0 and slope 1. If the assumption of normality is true, but the mean and the variance are not 0 and 1, the points still fit on a straight line, but the intersect and the slope of the line change to the mean and the standard deviation, respectively. If even normality is not satisfied, the points deviate systematically from the straight line. Figure \ref{fig:qq} shows the Q-Q plots of a constant star for all variants of scaling, with the expected standard normal line pictured in red, and the best-fitting normal distribution in blue. The sample is visibly not normal in any of the bands: the tails deviate from both lines, indicating heavier tails (the absolute values of the observations are larger  than what is expected from a normal distribution). The blue and the red lines have only slightly different slopes in most of the plots, suggesting that the scaling did obtain approximately unit variance, though the assumption of normality is not valid. 

This implies that the non-robust PCA, built on the assumption of normality, is not appropriate for our purposes. Another important issue that violates this distributional condition is the presence of outliers. We can see their effect in Fig. \ref{fig:pairplot}: a single outlier distorts completely the estimated normal distribution, outlined by the red ellipsoid. Robust versions of PCA are given in the statistical literature, among which we chose the minimum covariance determinant method \citep{rousseeuw85}. This involves repeated random subsampling of the data, leaving out a fixed percent of the observations each time. Then the covariance matrix is computed using each subsample, and the one having minimal determinant is chosen as the robust estimate. The blue ellipsoids in Fig. \ref{fig:pairplot} are the result of the application of this method. The percent of left-out data is fixed so that the effect of a single outlier is decreased, but that of a few consistently located outlying observations is preserved. This choice is motivated by two conflicting aims. One is to get rid off of the effects of true erroneous observations. The other is to preserve that of those that are true representatives of the light curve, though they are rare; for instance scarce measurements of the dips in detached eclipsing binaries. Indeed, the blue ellipsoids in Fig. \ref{fig:pairplot} represent better the apparent distribution of the bulk of the data.

Robust PCA yields some directly useful quantities. The first are the coefficients $\mathbf{v}_1$ of the linear combination, which characterize the direction of the first principal component. It depends on the type of variability, and thus can be used in classification. The second are the scores of the observations on the first principal component, that is, the time series of the linear combinations $\mathbf{z}_1 = \mathbf{X}\mathbf{v}_1$. By convention, the linear coefficients $\mathbf{v}_1$ are fixed so that the variance of the noise after transformation remains one, so the improvement on the signal-to-noise ratio is $\sum_{m = 1}^M v_{m1} A_m$, where $A_m$ is the signal amplitude in band $m$ after scaling. The third is the variance $d_1^2$ of $\mathbf{z}_1$, which is related to the elongation of the point cloud. Its ratio to the total variance $\sum_{m = 1}^M d_m^2$ (called hereafter the variance proportion) gives indication about the presence of variability. Finally, the distances of the observations from the centre with respect to the robust variance-covariance matrix fit indicate outlyingness of the observations, and can be used to weight the observations in period search.

\subsection{Period search} \label{subsec-periodsearch}

We used the generalised Lomb-Scargle method \citep{zechmeisterkurster09} to obtain the periodogram of the first principal component, and tested various trimming and weighting options on simulated sinusoids with errors imitating SDSS error patterns.

\begin{figure}
\begin{center}
\includegraphics[scale=.75]{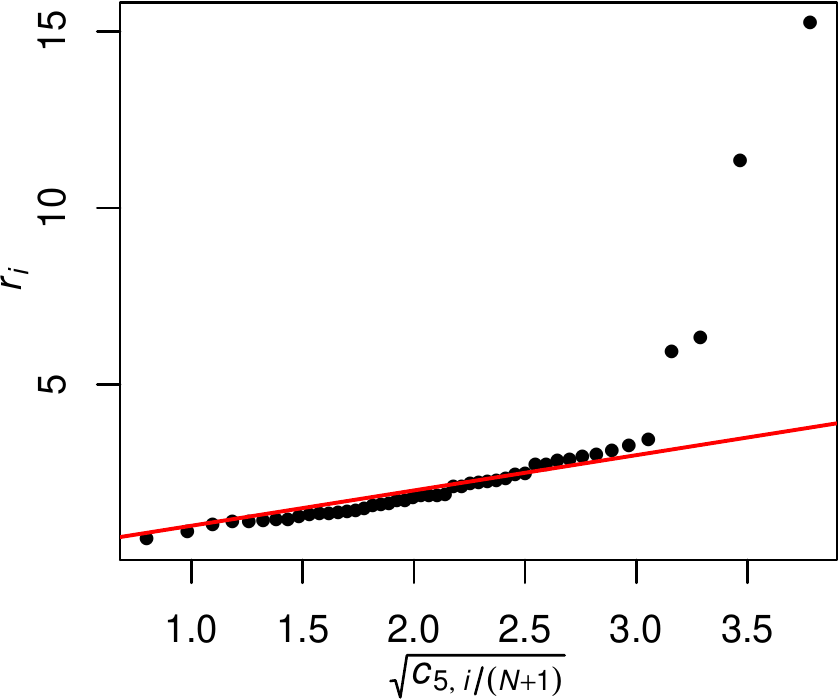}
\caption{ Robust distances from the center of the point cloud for the 5-dimensional observations of a star, on a square root scale for better visibility. The red line indicates the exact $\chi^2_5$ distribution.}
\label{fig:qqrobustdist}       
\end{center}
\end{figure}

The usual weighting with the inverse squared errors cannot be applied for the $\mathbf{z}_1$ time series, because the scaling and the construction of the principal components lead to approximate unit variance of the noise on $\mathbf{z}_1$. Instead, we can base a measure of outlyingness on the robust distances, and trim or weight the elements of the time series according to this measure. The squared distances, $r_{i}^2$, in an $M$-dimensional space under the hypothesis of approximate standard normality should follow a $\chi^2_M$ distribution. Similarly to the principle of Q-Q plots, we can determine what value is expected for the $i$-th largest distance, $r_{(i)}^2$, among a sample of size $N$ according to the $\chi^2_M$ distribution: this will be the $i/(N+1)$ quantile, $ c_{M, i/(N+1)}$, of the $\chi^2_M$ distribution ($N+1$ is taken instead of $N$ to avoid exactly 0 or 1 values). Such a $\chi^2_5$ Q-Q plot of the robust distances is shown in Fig. \ref{fig:qqrobustdist} for a variable star. We checked various alternatives of trimming thresholds and weight definitions based on the difference between the $i$th largest observed distance with simulations.

The goal of the simulations is to reproduce the SDSS error levels, error dependence on magnitude, outliers and observational cadences. In order to obtain this, we used the following procedure:
\begin{enumerate}[1.] 
\setlength{\itemsep}{-0.05cm}
\item From 2000 randomly selected objects from the 68,000 Stripe 82 objects flagged as variable, we extracted the sets of five amplitudes based on the excess root mean square variability, the sets of five median magnitudes and the $u$-band observational times. 

\item We generated 200 sine functions with random frequencies in the interval $[0,20]$ day$^{-1}$, with random phases, and randomly selected sets of five amplitudes and of median magnitudes from those extracted from our sample. Each sine function was multiplied by all five elements of one amplitude set, and one set of median magnitudes was added to it. We obtained thus 200 pure five-band light curves, each with coherent brightness variations in the bands. We sampled these sine waves with time cadences extracted from Stripe 82 in the previous step.

\item We added noise to the light curves. We divided the full magnitude range of the 2000 stars in each filter into 0.5 mag wide bins, and grouped the error bars according to the magnitude value with which it occurred. We obtained thus a sample of all error bars, in magnitude bins, for all bands. Then for each simulated magnitude values  of the sinusoidal light curves in each band,  we randomly selected an error bar value $\epsilon$ from the bin corresponding to the simulated magnitude value in that band. We generated a random number from the $\mathcal{N}(0, \epsilon^2)$ distribution, and added this value as the realised noise to the simulated magnitude. Finally, we joined $\epsilon$ as the error bar to the simulated series.

\item To simulate outliers, we changed some observations to fainter values. The differences were random values from $\mathcal{N}(0,(6\epsilon)^2)$, in a random number of filters simultaneously. The number of outliers followed a Poisson distribution with mean equal to  0.005.
\end{enumerate}
 
Visual inspection and Q-Q plots showed that the observed and simulated light curves are very similar. The simulations were then used to investigate the performance of different combinations of scaling, weighting or trimming and period search. We found that  the combination leading to the highest frequency recovery rate was the generalised Lomb-Scargle method with weights defined by
\begin{eqnarray} \label{eq_weights}
w_{i} = \frac{1}{W} \left[ \max \left\{1, \left| r_i^2 - c_{5, i/(N+1)} \right| \right\} \right]^{-1/2}
\end{eqnarray}
with $W = \sum_{i=1}^N  \left[ \max \left\{1, \left| r_i^2 - c_{5, i/(N+1)} \right| \right\} \right]^{-1/2}$, applied to either the $\mathbf{z}_1$ time series derived from variance stabilized observations or to observations scaled with robust estimators of the mean and standard deviation. 

 \begin{figure}
\begin{center}
\includegraphics[scale=.65]{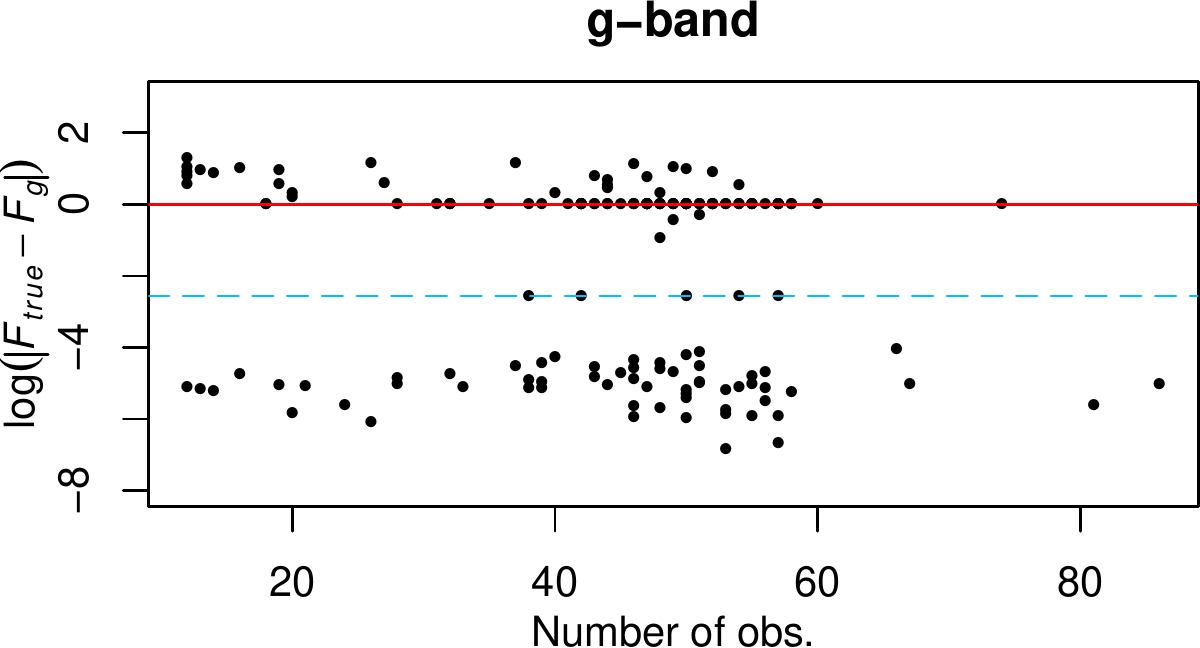}
\includegraphics[scale=.65]{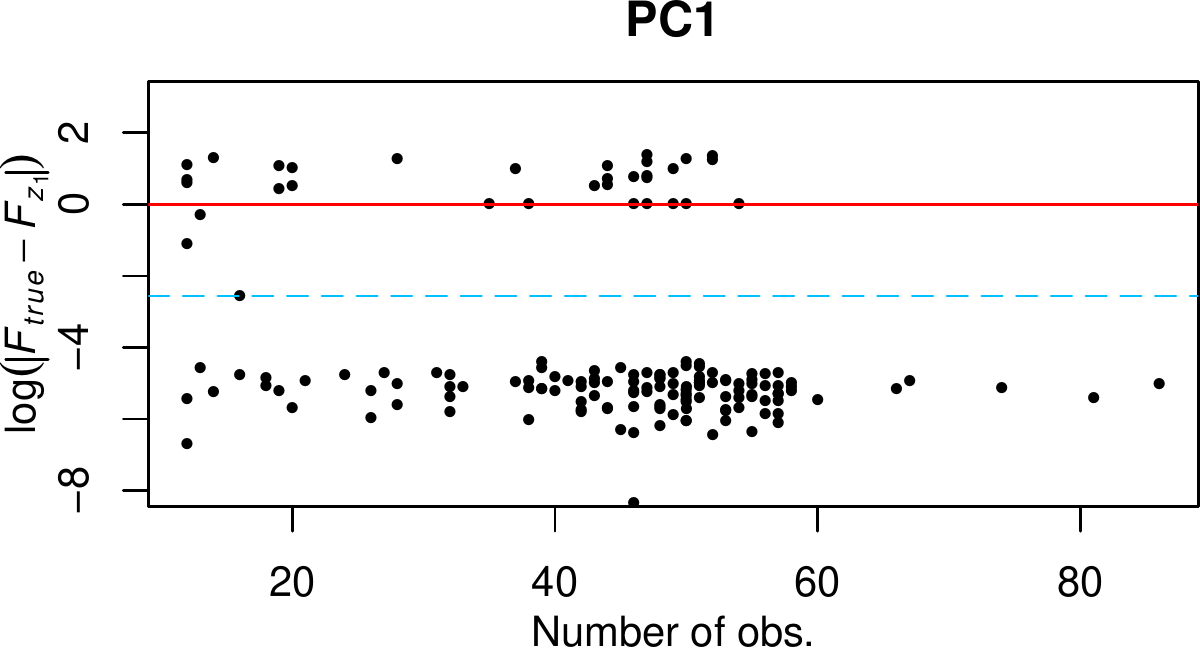}
\caption{The logarithm of the difference between the true and the found frequency in day$^{-1}$ for sinusoidal light curve simulations with approximate SDSS error distributions on $g$-band (top panel) and on PC1 (bottom panel). The solid red and dashed blue lines are the daily and the yearly aliases, respectively.}
\label{fig:psearch}       
\end{center}
\end{figure}

For comparison, the generalised Lomb-Scargle method with the usual weights based on error bars was also applied to detect periodicity on the single $g$ band, which generally had the best signal-to-noise ratio. The results of the best $\mathbf{z}_1$ and the single-band analysis are compared in Fig. \ref{fig:psearch}. The logarithm of the difference between the found and the true frequency is shown versus the number of observations in the time series. Period search on $\mathbf{z}_1$ outperforms the single-band analysis. While the single-band analysis finds a yearly or daily alias in 31 cases, the analysis of $\mathbf{z}_1$ reduces this number to only 8.

Based on these simulation results, in the analysis of the variable stars in Stripe 82 we applied the generalised Lomb-Scargle method with weights defined by \eqref{eq_weights}. For some variable stars, we also performed a multifrequency analysis. The folded $\mathbf{z}_1$ light curves were fitted with B-splines, restricting the smoothing parameter to provide a reasonable degree of smoothing (R procedure {\tt smooth.spline} with the constraint $0.5 \leq \mathtt{spar} \leq 1$ \citep{R}, and selecting its optimal value by leave-one-out cross-validation. The smoothed light curves $\hat{\mathbf{z}}_1$ were inspected visually, and the rare over- or undersmoothed cases were corrected by a manual selection of the smoothing parameter. The frequency analysis of the residuals was not performed on light curves  where the equivalent degrees of freedom of the smoothing was high with respect to the number of data points, since this left too few residual degrees of freedom to obtain meaningful harmonic fits in the residual frequency spectrum. For the cases with enough residual degrees of freedom, the residuals  $\mathbf{r} = \mathbf{z}_1 - \hat{\mathbf{z}}_1$ were extracted from the B-spline model, and the same period search algorithm with the same weights as for $\mathbf{z}_1$ was performed on their time series. The significance of eventual peaks in the residual periodogram  was assessed using a combination of non-parametric bootstrap \citep{davisonhinkley} and extreme-value methods \citep{coles}.

\subsection{Random Forest}\label{subsec_rf}
 
Supervised classification methods estimate the class (usually a discrete variable like $1, 2, \ldots, L$ or RRAB, RRC, EA, EB, ...) for an object of unknown class based on some attributes (for example period, amplitude, colour, etc.), by fitting a model to a set of objects with measured attributes and known classes (the training set), and then using this for prediction. Random Forest  \citep{breiman01, hastieetal} is a popular method which works excellently in a very wide range of data mining problems. It consists of building a collection (`forest') of classification trees on the training set, then passing any new instance down on all trees and obtaining the class estimate by the majority vote of the forest. 
 
 The training (`growing of the forest') begins with a selection of a large number of bootstrap samples from the  training set. Then a classification tree is built separately for each sample with binary splits. First a relatively small subset of attributes is selected randomly. For each of these, the split-point is computed that separates the given bootstrap sample with the least mixing of classes by some criterion (for example, one of the subsets contains only classes A and B, whereas the other mostly C and D). Then the tree is split according to the attribute for which this separation is the cleanest. For the next step, both subsets (called nodes) are further split in the same way as was done in the first step: from a small random subset of attributes, the one is  selected that splits best the node in question. The procedure is repeated until each final node is either perfectly clean or has a pre-defined minimal size. Then, using another bootstrap sample from the training set, a new tree is built. As a result,  a large forest of many trees emerges. To classify a new object, the prediction of the class by each tree in the forest is calculated, and the class with majority vote is accepted as the estimate. An estimate of the probabilities to belong to each of the classes can be obtained by calculating the proportion of votes for each class.
 
 The main advantage of this procedure is that it obtains the class estimate as an average of many individual estimates by the trees. These estimates are unbiased, but have high variance. Averaging preserves unbiasedness and reduces variance, and this variance reduction is larger, if the trees are less correlated \citep{hastieetal}. Random Forest achieves de-correlation by applying two tricks: first, it uses only bootstrap samples from the training sets, so the basic data set driving the learning process is not identical for each tree; and second, it uses the best of only a random subset of attributes, not of all attributes. The consequent higher variance of the individual trees is more than compensated by the decrease of the variance of the average because of the nearly non-correlated trees.
 
There are a few tuning parameters in the Random Forest procedure: the number of trees in the forest, the number of random attributes at the nodes, and the final node size. For most classification problems, growing several hundred trees is enough, and adding more trees does not improve the prediction accuracy. With respect to the final node size, optimal results are achieved by maximally growing the trees (that is, the minimal final node size is 1), but it is customary in larger problems with many instances and attributes to grow the trees only to a somewhat larger final node size to speed up the process without great loss of accuracy. The most influential tuning parameter is in general  the number of attributes from which the best split is chosen at the nodes, though Random Forest is only weakly sensitive to this as well. \citet{breiman01} proposes to choose the largest integer $k$ such that $k < \sqrt{K}$, where $K$ is the number of attributes.

\section{Application to Data} \label{sec_appl}

\subsection{Principal Component Analysis} \label{subsec_PCAdata}

We considered only stars that have at least 10 complete $ugriz$ observations from our data set. As discussed in Section \ref{subsec_PCA}, scaling and centring is necessary before applying robust PCA. We tested all four methods (local scaling, robust and non-robust average scaling and variance stabilization) outlined there for variability detection, characterization and period search. The requirements of these three tasks are not the same, and thus different scaling methods perform better in each one, but inspection of the results suggested that variance stabilization yields the best overall performance. We present here only the results based on this transformation. 

\subsubsection{Variability detection based on $\mathbf{z}_1$ variance} \label{subsubsec_vardet}

After scaling and centring the data, we expect to see a spherical point cloud, similar to a multivariate standard normal sample, if the errors are nearly non-correlated and the star is constant. Detection of variability is thus equivalent to detecting excess variance in the point cloud of the scaled observations as compared to the variance of the first principal component of a 5-dimensional standard normal point cloud.
Samples from even a spherically symmetric distribution show stochastic distortions from the perfect sphere, and the smaller the number $N$ of the points in the cloud is, the stronger this distortion is. As PCA selects the direction of maximal spread in the data, we cannot expect $d_1^2$ to be exactly 1. The distribution of  $d_1^2$ as a function of $N$ is the easiest to obtain with simulations. The zero hypothesis of non-variability of a source may be tested at a given confidence level $\alpha$ by comparing the observed $d_1^2$ of the source to the quantile $c_{1-\alpha}$ of the simulated distribution: $d_1^2 > c_{1-\alpha}$ rejects the hypothesis of non-variability of the source at the level $\alpha$. 

 \begin{figure}
\begin{center}
\includegraphics[scale=.75]{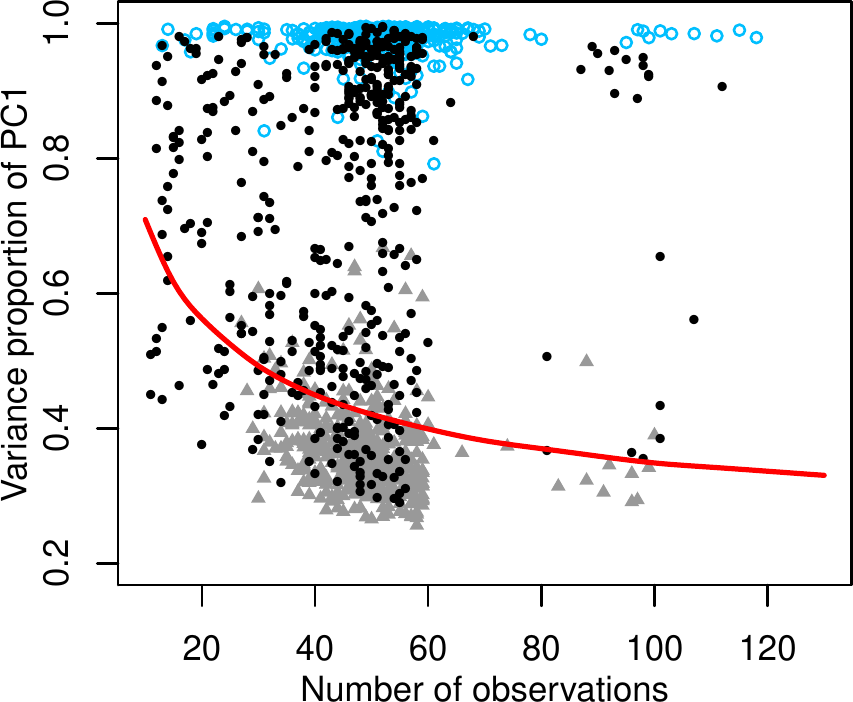}
\caption{The proportion of the variance of the first principal component to the total variance versus the number of observations for the RR Lyrae sample of \citet{sesaretal10} (empty blue circles), and for 500-500 variable and non-variable random objects (black dots and grey triangles, respectively). The simulated 0.9999 quantiles of the simulated distributions are added as a red line.}
\label{fig:vardetect}       
\end{center}
\end{figure}

As a PCA-based selection criterion, we use  the proportion of  $d_1^2$ to the total variance. Figure \ref{fig:vardetect} shows its comparison to the variable selection of  \citet{sesaretal10}. Five hundred objects that were flagged variable (black dots) and 500 flagged non-variable  (filled grey triangles) are plotted, together with the RR Lyrae sample of \citet{sesaretal10} (empty blue circles). The 0.9999 quantile derived from the simulations is denoted by the red line. For sources that are above this line, non-variability is rejected at the level of 0.0001. The left panel shows the proportion of  $d_1^2$ to the total variance. Most of the objects flagged variable (black dots) are above the red line, meaning that the variance proportion criterion and the classical cuts select approximately the same variable sample. Some difference nevertheless can be observed. Among the stars flagged non-variable, the PCA-based criterion found some variables: the grey triangles above the red line. The significant variance excess seems to be due to two reasons. When the bands are weakly correlated, the excess might be the consequence of either microvariability, or correlated and underestimated errors across bands \citep{scrantonetal05}, so to detect microvariability in Stripe 82, a de-correlated version of scaling should be used. The other is that the variance stabilization did not succeed to obtain unit variance in one band. In this case, the point cloud is elongated along one axis. Correspondingly, the coefficients $\mathbf{v}_1$ of the first principal component contain usually one value close to 1 at the band  which is the origin of the excess variance and a value close to 0 for all others. The characteristic profile of  $\mathbf{v}_1$ helps to recognize these cases.
 
Conversely, the black dots below the red line in the left panel of Fig. \ref{fig:vardetect} represent points that are flagged variable but found non-variable by PCA. These seem to belong to two groups: one for which the flag seems to originate from an outlier, and another that appears ball-like on the pairwise scatterplots like Fig.  \ref{fig:pairplot}, but not with unit variance. The variable flag from the classical analysis in the latter case can be due to under-estimated but nearly non-correlated errors in all bands \citet{scrantonetal05}. If the under-estimation is of similar order in every band, then the PCA-based criterion is less biased than traditional cuts, as it is defined as a proportion. Moreover, $\mathbf{v}_1$ would again contain one value close to 1 and all the others near-zero, so it is possible to filter out these cases based on the $\mathbf{v}_1$ profile.
In general, this form of $\mathbf{v}_1$ can be used to recognize and filter the cases where the excess scatter is due to underestimated errors in one or a few bands rather than true variability of the source. 

\subsubsection{Period search on PC1 time series} \label{subsubsec_psearch}

For the majority of the investigated variable objects, period search on the SDSS time series suffers from aliasing problems. The daily and yearly observational patterns (see e.g. \citealt{sesaretal07,sesaretal10}) create aliases with complex structures, displaying combinations of the variability frequency with the 1/day and 1/year frequencies with comparable amplitudes. According to the simulation results of Section \ref{subsec-periodsearch}, using the time series of  $\mathbf{z}_1$ with weights defined by \eqref{eq_weights} improves on period search on $g$-band. We applied this procedure, complemented by a non-linear optimization to find the exact value of the frequency, to a small random sample of 2000 stars flagged variable, which contained 36 known RR Lyraes. The published periods of these RR Lyraes, as described in \citet{sesaretal10}, were determined by a visual inspection of light curves folded with the five best periods returned by the SuperSmoother algorithm  \citep{reimann94} restricted to the [0.2, 1] day interval. For 30 out of the 36 RR Lyrae, the result from our algorithm coincided with the true frequency. In the 6 other cases, daily alias periods were found. This suggests that we can expect good period search results from a simple automatic  procedure even in the absence of visual inspection. 

We applied the weighted period search method (without the nonlinear optimization) on the full variable star set. The dominant frequency was used as an attribute in the classification procedures, presented in Section  \ref{subsec_rfappl}. The classification selected a sample of 317 candidate RR Lyrae and HADS stars.  For these, we modelled the folded $\mathbf{z}_1$ light curve, computed the residuals $\mathbf{r} = \mathbf{z}_1 - \mathbf{\hat z}_1$, and performed another period search on $\mathbf{r}$ as described in Section  \ref{subsec-periodsearch}. The found likely multi-period objects are discussed in  Sections  \ref{subsec_candrr} and \ref{subsec_candsx}.

\subsubsection{Characterisation of variability with PC1 spectrum} \label{subsubsec_charac}

 \begin{figure*}
\begin{center}
\includegraphics[scale=.84]{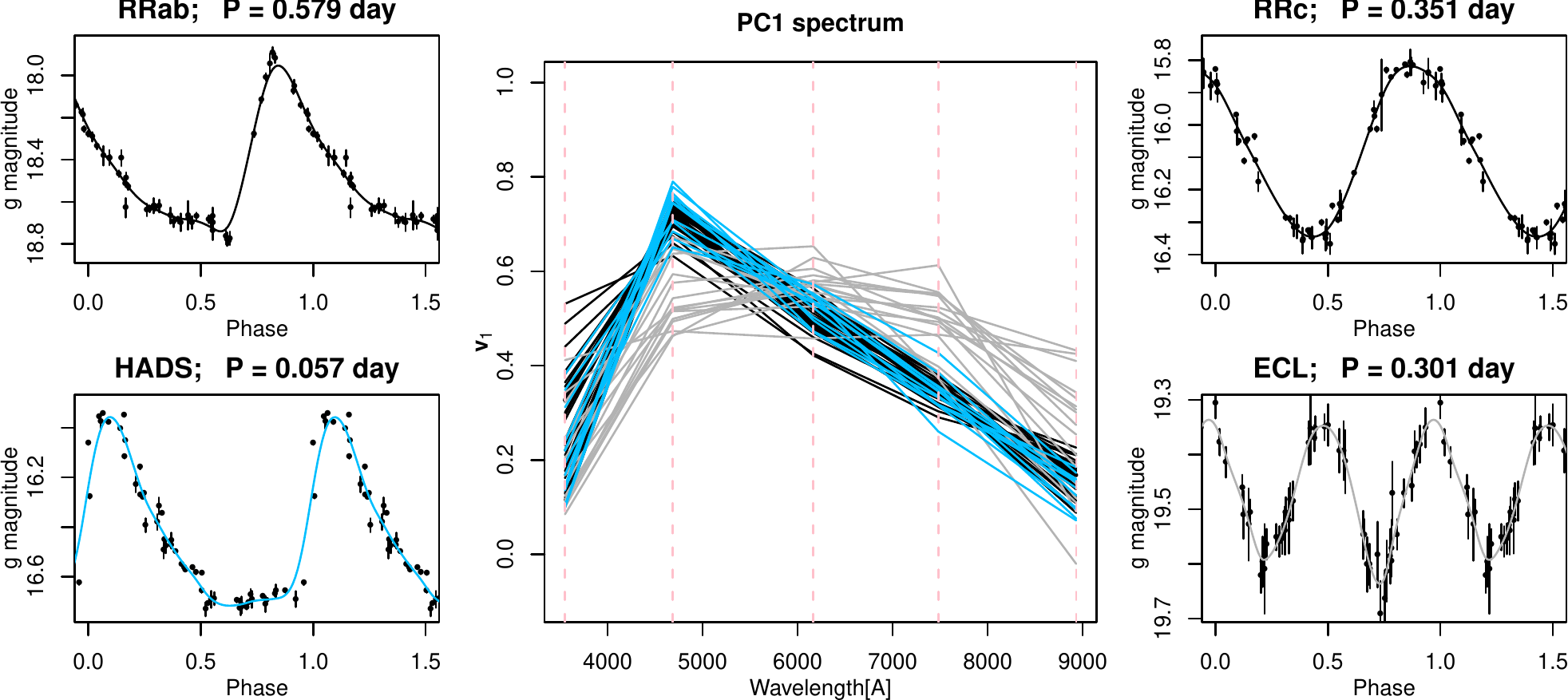}
\caption{Examples of folded light curves (left top: RRab, left bottom: HADS, right top: RRc, right bottom: eclipsing binary) and some PC1 spectra ($\mathbf{v}_1$) of the classes RR Lyrae (black lines), HADS (blue) and eclipsing binary (grey).  There is a marked difference between the two pulsating types and the eclipsing binaries. }
\label{fig:pcsp}       
\end{center}
\end{figure*}

RR Lyrae variables show a characteristic  wavelength-dependent amplitude pattern: their variability is stronger at shorter optical wavelengths than at the red end of the spectrum \citep{smith}. If we could observe RR Lyrae stars with SDSS filters with equal errors in all bands, we would see the first principal direction tilted  towards the blue rather than the red bands, and correspondingly, larger coefficients $\mathbf{v}_1$ for blue than for red wavelengths. However, the error patterns distort this simple picture. We must scale the bands in order to get rid off of the effect of different average error levels, and the downscaling will be stronger in bands with larger errors, most notably in $u$ and $z$. As a result, variability amplitudes get downscaled as well. Thus, the typical RR Lyrae  $\mathbf{v}_1$ pattern (the PC1 spectrum) takes a characteristic shape. It is composed of, on the one hand, the amplitude pattern of pulsating variables governed by the physical parameters and pulsation mode, and on the other, of the scaling patterns of the survey. Typical PC1 spectra of identified RRab and RRc stars are shown with black lines in the middle panel of Fig. \ref{fig:pcsp}. They exhibit small coefficients $v_{u1}$ and $v_{z1}$ for the $u$- and $z$-band contributions, and the highest value is $v_{g1}$ at the $g$-band, which has high variability amplitude and low errors. 

HADS variables are in some aspects similar to the RR Lyraes (see e.g. \citet{breger80, mcnamara95, petersenchristensendalgaard96, mcnamara97}; a light curve is shown in the bottom left panel of Fig. \ref{fig:pcsp}). Their effective temperature is roughly in the same range, and although their pulsation may be more complex than that of RR Lyraes, they are mainly radial-mode pulsators. They show similar patterns across the wavelengths with decreasing amplitudes from blue to red wavelengths \citep{pigulskietal06}, so we can expect their PC1 spectrum to be similar to that of the RR Lyraes. The middle panel of Fig. \ref{fig:pcsp} presents the PC1 spectrum of several variables that were found to have periods, colours and folded light curves characteristic to HADS  variables. These are plotted in blue, superposed on the lines of the RR Lyrae stars.

For some other types of variability, we can expect different PC1 spectra. Eclipsing binaries that are composed of two  components of the same age and similar masses have similar colours, and therefore show only very weak colour changes and an almost-equal contribution of all bands to the light variation. This results in a flatter PC1 spectrum. Combined with the specific error pattern of SDSS, this yields a profile that is low at the noisy $u$ and $z$ bands and have a higher plateau at $g$, $r$ and $i$. Several PC1 spectra of this type are also shown in the middle panel of Fig. \ref{fig:pcsp}; these objects have clear eclipsing binary-type folded light curves.

Figure \ref{fig:pcsp} shows the use of the PC1 spectrum for discriminating certain types. In the two panels on the right-hand side, we show the light curve of an RRc star from the confirmed sample of \citet{sesaretal10} (upper right panel) and an eclipsing binary candidate with only slightly different minima and near-sinusoidal light variation (bottom right). This eclipsing binary, when folded by half of the period, shows a folded light curve very similar to that of an RRc variable, as the tiny difference in the depths of the minima is masked by the error bars. If such eclipsing binaries fall furthermore close to the colour-colour region of RRc stars, the automated classification is difficult. However, the PC1 spectrum is approximately flat-topped for a geometric origin, whereas it has a peak at $g$ for a pulsating RRc. The inclusion of the first principal direction $\mathbf{v}_1$ can thus be helpful in an automated classification procedure.

\subsection{Random Forest classification} \label{subsec_rfappl}

\subsubsection{Preliminary  selection} \label{subsec_presel}

 \begin{figure*}
\begin{center}
\includegraphics[scale=.75]{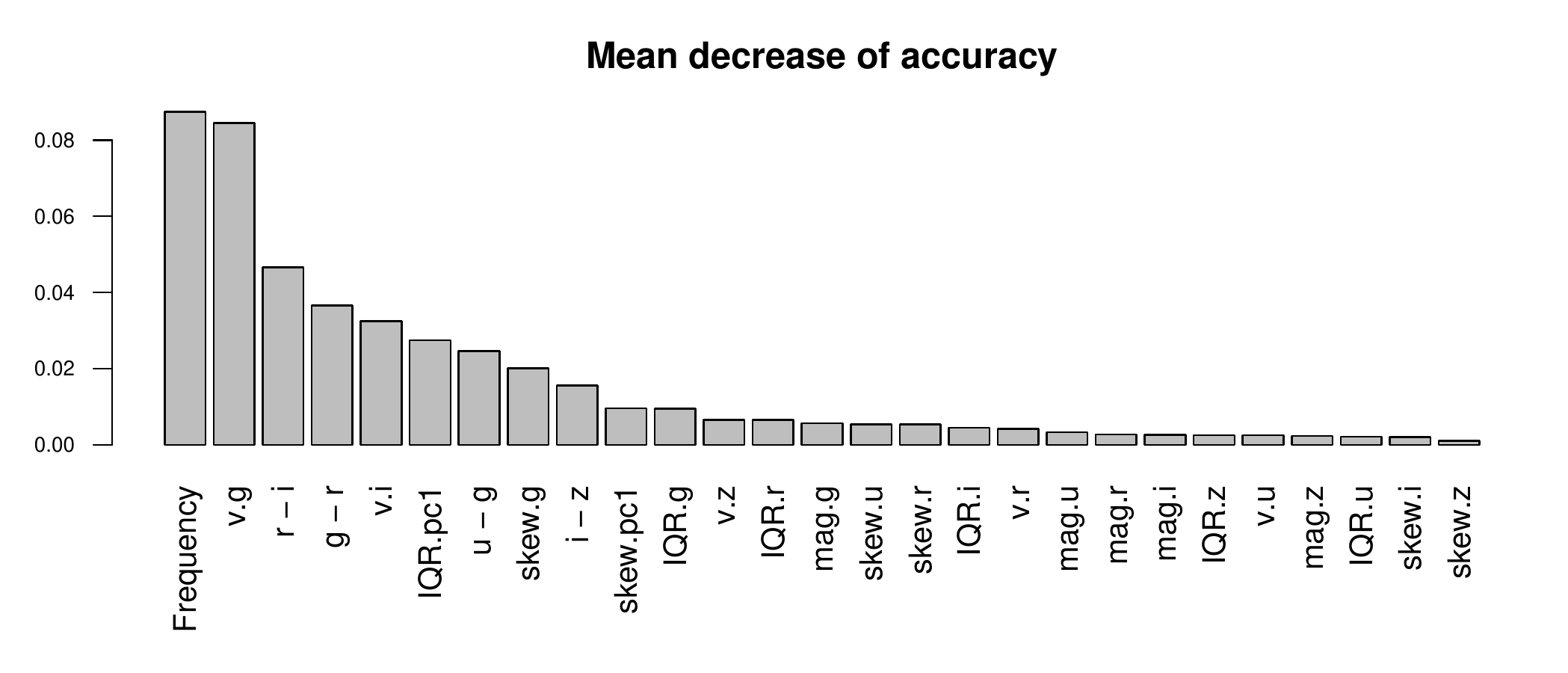}
\caption{Attribute ranking by mean accuracy decrease when classifying without the attribute in question. The components $v_{1g}$ and $v_{1i}$ of the PC1 spectrum are highly ranked, comparable to the frequency, $r-i$ and $g-r$ colours. 
}
\label{fig:attrank}       
\end{center}
\end{figure*}

As our goal is to extract only a few variable types of interest from the SDSS Stripe 82 sample, we define our classes as RRAB, RRC, HADS and O, corresponding to the four types RRab, RRc, HADS and `other'. Supervised classification needs a set of objects with known types and well-measured attributes to train the classifier. For RRab and RRc type variables, the sample of \citet{sesaretal10} provides such a set. For HADS stars, we do not have a confirmed database. However, they form a relatively well-defined, separable class of variable stars, so we can select visually a sample of plausible HADS candidates for the training set. Class O, on the other hand, is broad enough to elude all concise, easy-to-implement definition. Our only requirement can be precisely this broadness: class O in the training set should contain representatives of all alternative types of variables mixed together, except for RR Lyrae or HADS stars. A reasonable criterion is therefore the separability of this class from HADS, RRAB and RRC. Thus, to obtain a clean training set, we apply Random Forest classification on the visually selected sample, then we remove all HADS and O objects that were misclassified, and we iterate these steps, until we reach a sample in which these two classes are recognized cleanly. As the RRAB and RRC classes are already confirmed, we removed only the O objects that were classified as RRAB or RRC, but not the RRAB and RRC stars that were misclassified as O.

For the visual selection of HADS training sample, we have expressed all possible selection criteria in the framework of PCA. The extraction of variable objects was performed with the aid of the variance proportion of the first principal component. To reflect the astrophysical properties of the sought stars, we require them also to show the characteristic colour changes during the cycle. In the language of PCA, this corresponds to have broadly the PC1 spectrum demonstrated by the blue lines in the middle panel of Fig. \ref{fig:pcsp}. Furthermore, they must have admissible $u-g$ and $g-r$ colours. Also, the dominant frequency $F_{\mathbf{z}_1}$ of the time series of the first principal component $\mathbf{z}_1$ must correspond to the HADS frequency range. In summary, the selection criteria are as follows.
\begin{enumerate}[(i)]
\item PC1 has a variance proportion higher than the 0.9999-quantile of that of a 5-variate standard normal point cloud with the same number of observations.
\item  The PC1 spectrum $\mathbf{v}_1 = (v_{u1}, v_{g1},v_{r1},v_{i1},  v_{z1})^T $ satisfies the following cuts: 
$ 0 < v_{u1} < 0.8, \;\;  0.45 < v_{g1} \leq 1,
\;\;    0.3 < v_{r1} < 0.75,
\;\;    0.15 < v_{i1} < 0.6,
\;\;    0 < v_{z1} < 0.5.$
\item The (extinction-corrected) median colours are in the region  $ 0.7 < u-g < 1.4$ and $-0.2 < g-r < 0.4$ of the $u-g, \;\; g-r$ diagram.
\item  $F_{\mathbf{z}_1} > 3/$day.
\end{enumerate}

All cuts are given very broadly, so the resulting selection contains a majority of contaminating objects such as eclipsing binaries with flat PC1 spectra, quasars, main-sequence variables and even RRc stars. Visual inspection of the $g$-band light curves resulted in 117 plausible HADS candidates. The training set for the first cleaning iteration consisted of these as class HADS; the RRab and RRc variables of \citet{sesaretal10} as classes RRAB (379 objects) and RRC (104 objects), respectively; and a random selection of 2400 other (O) variables from the remaining variable set, yielding a training set of 3000 objects in total. For all objects we calculated the following attribute list:
\begin{itemize}
\item the median (apparent) magnitudes in all bands, corrected for interstellar extinction using the dust map of \citet{schlegeletalxx};
\item the median (extinction-corrected) colours $u-g$, $g-r$, $r-i$ and $i-z$;
\item  the interquartile range IQR$_u = q_{u}(0.75) -  q_{u}(0.25), \ldots$ of every band and of the first principal component scores $z_{11}, \ldots, z_{1T}$ as a percentile-based estimate of the light curve range, where $q_{x}(p)$ denotes the empirical $p$-quantile of the time series $x$ ($x = u,g,r,i,z$ and PC1);
\item percentile estimate $S_u = {[q_{u}(0.9) + q_{u}(0.1) - q_{u}(0.5)]}  $ $[q_{u}(0.9) - q_{u}(0.1)]^{-1}, \ldots$ of the light curve skewness for all bands, in addition to that of the first principal component scores;
\item the PC1 spectrum $\mathbf{v}_1 = (v_{u1}, v_{g1},v_{r1},v_{i1},  v_{z1})^T $;
\item the dominant frequency $F_{\mathbf{z}_1}$ found by the period search method described in Section \ref{subsec-periodsearch}. 
\end{itemize}

Selection of a small but sufficient attribute set is nevertheless necessary: the performance of most data mining procedures is sensitive to the presence of noise-like attributes. The importance of any attribute can be measured by Random Forest by for example calculating the average accuracy loss on trees generated without the use of the attribute in question. The importance plot for all attributes in Fig. \ref{fig:attrank} indicates that the best two attributes are the frequency and the $g$-band component $v_{g1}$ of the first principal direction. They are followed by the $r-i$ and $g-r$ median colours, then another two PCA-based attributes, $v_{i1}$ and the IQR of the ${\mathbf{z}_1}$ time series. This latter proves to be more important than the $g$-band IQR. The relatively low accuracy loss due to omission of colours is a consequence of our preselection of colour range in the $u-g, \; g-r$ plane and of the correlation between colours. According to a procedure similar to that applied in Sections 4.2-4.4 of \citet{dubathetal11}, the most performant attribute set is the 11 top-ranking attributes: $ F_{\mathbf{z}_1},  v_{g1}, r-i, g-r,  v_{i1}, \mathrm{IQR}_{\mathbf{z}_1}, u-g, S_g, i-z, S_{\mathbf{z}_1}$ and $\mathrm{IQR}_{g}$. This is what we used for the following cleaning of the training set and the extraction of our  set of candidate RR Lyrae and HADS stars.

The next stage is the cleaning of the training set, to obtain clearly separated HADS and O classes. In the first iterative step of this, Random Forest was trained on the 3000 objects using the attribute list presented above. At this stage, the set O may contain unrecognized RR Lyrae and HADS stars. We assume that this contamination is low enough not to bias strongly the automated classification results for class O in the first run, as these are relatively rare compared to all other types taken together. An overall error rate less than 1\%  in the first iteration confirmed this assumption. After each run of Random Forest, we removed the objects from type O and HADS that were misclassified (keeping all RR Lyrae regardless of predicted type, since these are confirmed variables), and repeated classification. After 3 iterations, we ended up with 108 HADS and 2372 O stars, which were separated perfectly from each other and from the RR Lyrae classes. The only confusion arose from 10 RRAB and RRC stars that were classified as type O, and one RRAB classified as RRC. This procedure biases the subsequent extraction of HADS, RRAB and RRC towards purity against completeness, as we dropped all cases that might represent unusual specimens of classes, and the final selection does not fully take into account eventual real similarities between stars belonging to distinct types. The choice between purity and completeness is certainly subjective, and depends on the purpose of the study. 

Re-processing the whole variable collection by Random Forest using the resulting training set gave 317 candidates, 163 HADS, 82 RRC and 72 RRAB stars. Multifrequency analysis as described in Section \ref{subsec-periodsearch} was performed on them. The final visual check was done with the help of  `portraits' of the stars: summary plots (Figures \ref{fig:rrab}--\ref{fig:doublemodeSX}) that contained the most important information that could have been extracted from the data, namely,  $\mathbf{z}_1$ and residual frequency spectra, folded $\mathbf{z}_1$, $g$, $g-i$ and residual light curves, the raw observed time series, colour-colour, colour-period and period-amplitude diagrams, and PC1 spectrum. These plots are presented in the Appendix. The visually selected 132 HADS, 68 RRab, 36 RRc and 25 multiperiodic or peculiar RR candidates are discussed in the next two sections.


\subsubsection{Candidate RR Lyrae variables} \label{subsec_candrr}


\begin{table*}
{\scriptsize
\caption{RR Lyrae candidates in Stripe 82. ID: object identifier; R.A.: right ascension (deg); Decl.: declination (deg); $u_{med}$,...: median magnitudes corrected for the ISM extinction;  $F_{\mathbf{z}_1}$: pulsation frequency determined from generalised least squares period search on PC1; $A_u$,...: amplitudes from a B-splines fit to the folded light curves. The complete table is available online.}
\begin{tabular}{ccccccccccccccc}
  \hline
ID & R.A. & Decl. & Type & $u_{med}$ & $g_{med}$ & $r_{med}$ & $i_{med}$ & $z_{med}$ & $F_{\mathbf{z}_1}$ & $A_{u}$ & $A_{g}$ & $A_{r}$ & $A_{i}$ & $A_{z}$ \\ 
  \hline
510582 & -11.901668 & -0.947130 & RRAB & 20.22 & 19.14 & 18.90 & 18.79 & 18.79 & 1.24630 & 0.20 & 0.19 & 0.13 & 0.12 & 0.10 \\ 
  651051 & 19.017164 & 1.262702 & RRAB & 18.27 & 17.07 & 16.80 & 16.72 & 16.69 & 1.77625 & 0.84 & 0.97 & 0.63 & 0.50 & 0.48 \\ 
  1013845 & 23.603725 & -0.592160 & RRAB & 17.32 & 16.11 & 15.87 & 15.77 & 15.77 & 1.36722 & 0.78 & 0.75 & 0.67 & 0.55 & 0.43 \\ 
  1918041 & -31.227027 & -0.703085 & RRC & 20.83 & 19.69 & 19.62 & 19.61 & 19.69 & 2.65171 & 0.34 & 0.44 & 0.26 & 0.20 & 0.19 \\ 
  2654711 & -41.834262 & -0.994213 & RRC & 17.66 & 16.48 & 16.54 & 16.62 & 16.69 & 4.88670 & 0.16 & 0.19 & 0.14 & 0.11 & 0.10 \\ 
  2725572 & -40.845223 & -0.719055 & RRAB & 20.15 & 19.03 & 18.74 & 18.69 & 18.69 & 1.77540 & 1.40 & 0.31 & 0.71 & 0.63 & 0.49 \\ 
  3352291 & -46.471817 & 1.260386 & RRAB & 15.85 & 14.67 & 14.41 & 14.33 & 14.34 & 1.95774 & 1.18 & 1.20 & 0.86 & 0.69 & 0.63 \\ 
  \vdots &  &  &  &  &  &  &  &  &  &  &  &  &  &  \\ 
   \hline
\end{tabular}
\label{table:rrall}
}
\end{table*}

\begin{table*}
{\scriptsize
\caption{Best-fitting templates of \citet{sesaretal10} for the new RR Lyrae candidates in Stripe 82. ID: object identifier; R.A.: right ascension (deg); Decl.: declination (deg); $d$: heliocentric distance for stars included in Fig. \ref{fig:RRmap};  $\langle V \rangle$: mean dereddened $V$-band magnitude, determined from a best-fit $V$-band template synthetized from the best-fitting $g$ and $g$ templates, $A'_u, ...$: amplitudes; $\phi_0^u$: epoch of maximum brightness; $u_0$: the magnitude at the epoch of maximum brightness corrected for the ISM extinction (the last three determined from the best-fit template);  $T_u,...$: best-fit template identifier number as given in \citet{sesaretal10}. The asterisks indicate slightly underestimated values. $d$ and $\langle V \rangle$ were computed only for those RRab variables that are plotted on Fig. \ref{fig:RRmap}, and are therefore missing for some stars. The complete table is available online.}
\begin{tabular}{cccccccccccc}
  \hline
ID & R.A. & Decl. & $d$ & $\langle V \rangle$ & Period & $A'_u$ & $\phi_0^u$ & $u_0$ & $T_u$  & $A'_g$ & $\ldots$ \\ 
  \hline
510582 & -11.901668 & -0.947130 &  &  & 0.802373 & 0.194 & 54358.25 & 20.113 & 104 & 0.187 & $\ldots$ \\ 
  651051 & 19.017164 & 1.262702 & 17.256 & 16.785 & 0.562973 & 0.604* & 53639.39 & 17.766 & 100 & 0.663* & $\ldots$ \\ 
  1013845 & 23.603725 & -0.592160 & 11.322 & 15.870 & 0.731397 & 0.687* & 54029.29 & 16.816 & 101 & 0.659* & $\ldots$ \\ 
  1918041 & -31.227027 & -0.703085 &  &  & 0.377115 & 0.375 & 54012.05 & 20.612 & 0 & 0.416 & $\ldots$ \\ 
  2654711 & -41.834262 & -0.994213 &  &  & 0.204637 & 0.161 & 53675.09 & 17.567 & 0 & 0.167 & $\ldots$ \\ 
  2725572 & 319.154777 & -0.719055 & 42.543 & 18.744 & 0.563272 & 0.626 & 53625.21 & 19.650 & 102 & 0.666* & $\ldots$ \\ 
  3352291 & 313.528183 & 1.260386 & 5.600 & 14.341 & 0.510786 & 0.559* & 53704.09 & 15.349 & 100 & 0.598* & $\ldots$ \\ 
  \vdots &  &  &  &  &  &  &  &  &  &  &    \\ 
   \hline
\end{tabular}
\label{table:rrtempl}
} 
\end{table*}

 \begin{figure}
\begin{center}
\includegraphics[scale=.49]{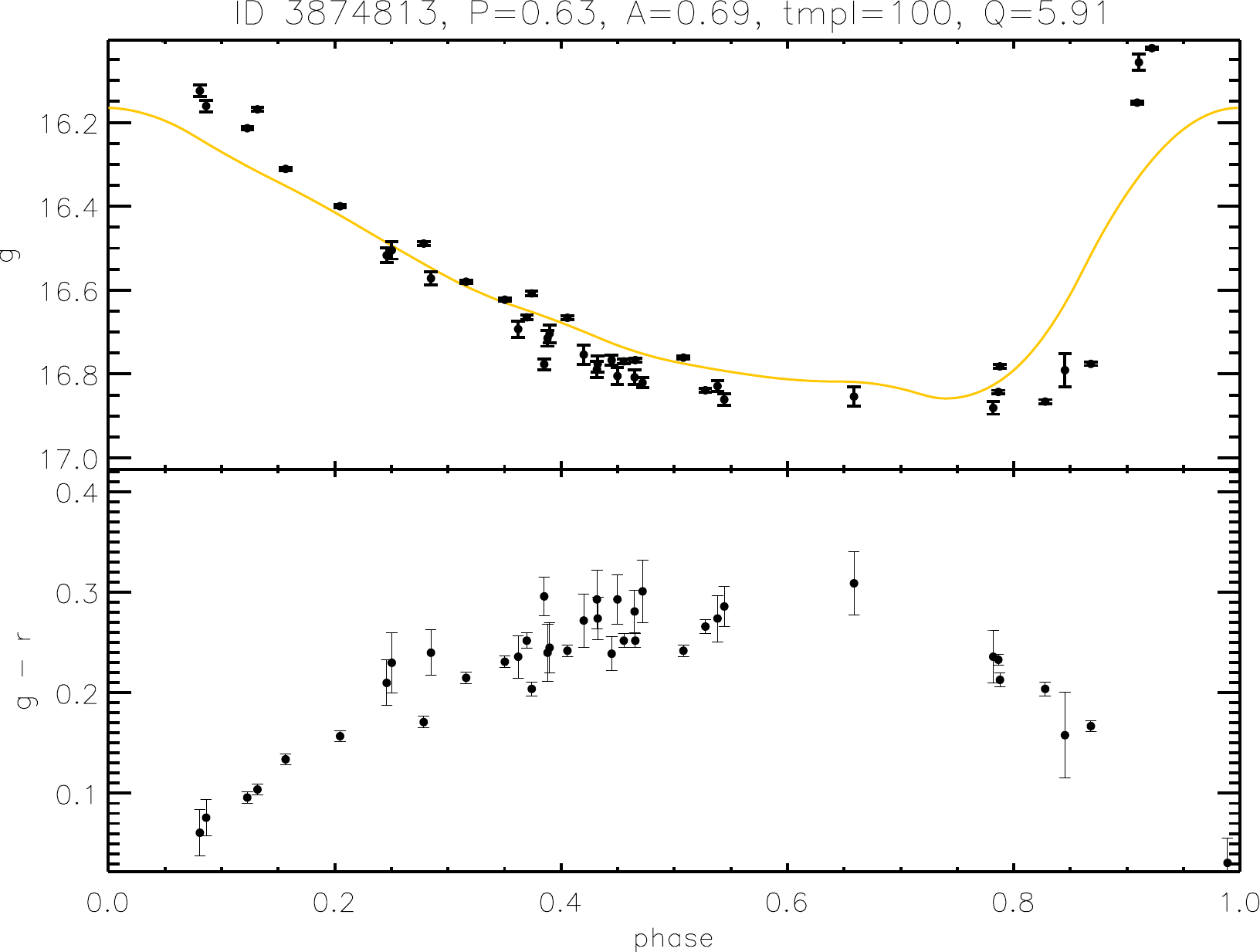}
\caption{$g$-band template fit for star 3874813, using the templates of \citet{sesaretal10}, an example that cannot be fitted well by these. For comparison, see the second panel from top in the middle column of Fig. \ref{fig:rrab}, which shows the same star fitted with a B-spline.
}
\label{fig:branifit}       
\end{center}
\end{figure}

\begin{figure}
\begin{center}
\includegraphics[scale=.68]{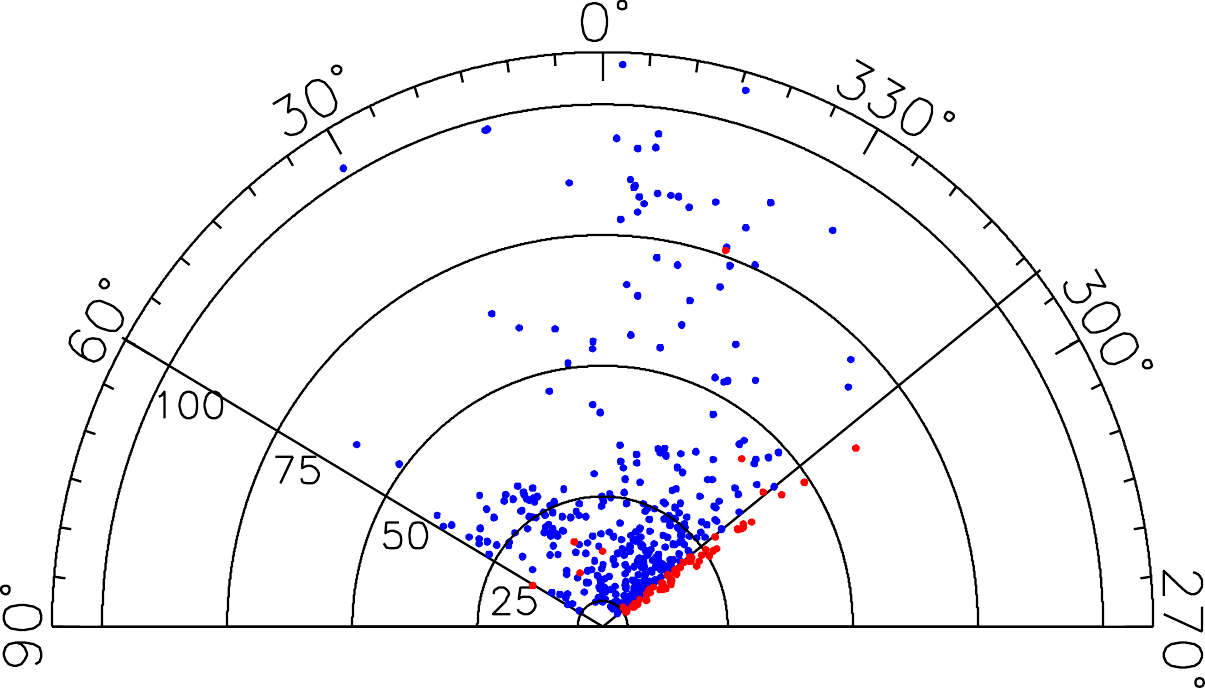}
\caption{Heliocentric map of the Stripe 82 RRab stars, based on average halo metallicity value.  The radial axis is the heliocentric distance in kpc, the angle is the equatorial right ascension. Blue dots: the sample of \citet{sesaretal10}, red dots: the new candidate RRab.
}
\label{fig:RRmap}       
\end{center}
\end{figure}

\begin{table*}
{\scriptsize
\caption{Double-mode RR Lyrae candidates in Stripe 82. Notation is the same as for Table \ref{table:rrall}. $F_{\mathbf{r}}$: secondary frequency, Ratio: ratio of fundamental to first overtone frequency, $P_{\mathbf{r}}$: P-value of secondary frequency. }
\begin{tabular}{ccccccccccccccccc}
  \hline
ID & R.A. & Decl. & $u_{med}$ & $g_{med}$ & $r_{med}$ & $i_{med}$ & $z_{med}$ & $F_{\mathbf{z}_1}$ & $F_{\mathbf{r}}$ & Ratio & $P_{\mathbf{r}}$ & $A_{u}$ & $A_{g}$ & $A_{r}$ & $A_{i}$ & $A_{z}$ \\ 
  \hline
1059995 & -23.605640 & -0.056749 & 21.02 & 19.94 & 19.75 & 19.75 & 19.78 & 2.48876 & 1.85620 & 0.74583 & 0.000 & 0.19 & 0.36 & 0.27 & 0.22 & 0.09 \\ 
  1346981 & 25.772163 & 1.097024 & 18.20 & 17.09 & 17.03 & 17.03 & 17.04 & 2.82707 & 2.10280 & 0.74381 & 0.000 & 0.40 & 0.43 & 0.31 & 0.23 & 0.20 \\ 
  1638185 & 30.812051 & 1.205761 & 18.05 & 16.93 & 16.80 & 16.80 & 16.80 & 2.84693 & 2.11660 & 0.74347 & 0.000 & 0.48 & 0.49 & 0.37 & 0.31 & 0.24 \\ 
  1875049 & -30.900196 & 0.941754 & 19.66 & 18.57 & 18.50 & 18.48 & 18.48 & 2.41850 & 1.80320 & 0.74558 & 0.000 & 0.20 & 0.35 & 0.24 & 0.19 & 0.22 \\ 
  2249641 & -39.804740 & 0.210125 & 17.47 & 16.38 & 16.31 & 16.32 & 16.34 & 2.78830 & 2.07536 & 0.74431 & 0.000 & 0.45 & 0.48 & 0.33 & 0.26 & 0.22 \\ 
  2662389 & -44.787449 & 0.404872 & 18.22 & 17.13 & 17.06 & 17.08 & 17.11 & 2.84001 & 2.10780 & 0.74218 & 0.000 & 0.38 & 0.44 & 0.30 & 0.24 & 0.21 \\ 
  2740388 & -44.211467 & -1.202920 & 17.61 & 16.54 & 16.47 & 16.45 & 16.50 & 2.76911 & 2.06080 & 0.74421 & 0.000 & 0.45 & 0.49 & 0.34 & 0.28 & 0.21 \\ 
  3091639 & 48.387957 & 0.715216 & 19.33 & 18.24 & 18.14 & 18.16 & 18.19 & 2.82369 & 2.09916 & 0.74341 & 0.000 & 0.44 & 0.55 & 0.36 & 0.29 & 0.23 \\ 
  3182847 & -48.344716 & 0.330749 & 20.53 & 19.42 & 19.33 & 19.36 & 19.40 & 2.77071 & 2.06144 & 0.74401 & 0.000 & 0.38 & 0.50 & 0.39 & 0.28 & 0.29 \\ 
  3524879 & -49.385512 & -0.480494 & 17.52 & 16.37 & 16.28 & 16.27 & 16.29 & 2.42901 & 1.81232 & 0.74612 & 0.000 & 0.39 & 0.47 & 0.32 & 0.26 & 0.21 \\ 
  4185977 & -52.885267 & -0.892139 & 15.96 & 14.84 & 14.71 & 14.70 & 14.75 & 2.64865 & 1.97572 & 0.74593 & 0.000 & 0.48 & 0.53 & 0.38 & 0.30 & 0.26 \\ 
  5631911 & -57.845165 & -0.839656 & 19.85 & 18.74 & 18.66 & 18.57 & 18.63 & 2.43680 & 1.81908 & 0.74650 & 0.000 & 0.49 & 0.51 & 0.34 & 0.27 & 0.26 \\ 
  1149344 & -21.758206 & 0.077395 & 20.22 & 19.16 & 19.03 & 18.99 & 18.99 & 2.40371 & 2.79388 & 0.86035 & 0.000 & 0.32 & 0.37 & 0.25 & 0.18 & 0.18 \\ 
  1283137 & 27.742275 & -0.847584 & 19.08 & 17.92 & 17.83 & 17.82 & 17.84 & 1.75063 & 2.04704 & 0.85520 & 0.000 & 0.40 & 0.48 & 0.32 & 0.24 & 0.22 \\ 
   \hline
\end{tabular}
\label{table:rrd}
}
\end{table*}

\vspace{6mm}
\begin{table*}
{\scriptsize
\caption{ RR Lyrae candidates with closely spaced multiple frequencies or indicating period, phase or amplitude  shifts in Stripe 82.  Notation is the same as for Table \ref{table:rrd}.}
\begin{tabular}{ccccccccccccccccc}
  \hline
ID & R.A. & Decl. & $u_{med}$ & $g_{med}$ & $r_{med}$ & $i_{med}$ & $z_{med}$ & $F_{\mathbf{z}_1}$ & $F_{\mathbf{r}}$ & Ratio & $P_{\mathbf{r}}$ & $A_{u}$ & $A_{g}$ & $A_{r}$ & $A_{i}$ & $A_{z}$ \\ 
  \hline
1139404 & -24.140121 & 0.369111 & 17.71 & 16.54 & 16.54 & 16.57 & 16.62 & 3.63296 & 3.62848 & 0.99877 & 0.000 & 0.33 & 0.38 & 0.28 & 0.21 & 0.18 \\ 
  1945634 & -31.463604 & 1.030967 & 17.50 & 16.32 & 16.11 & 16.01 & 15.98 & 1.53242 & 1.53904 & 0.99570 & 0.035 & 0.52 & 0.58 & 0.37 & 0.27 & 0.26 \\ 
  2954798 & -41.387218 & -0.569949 & 18.15 & 16.95 & 16.97 & 17.01 & 17.06 & 3.15265 & 3.15188 & 0.99976 & 0.001 & 0.46 & 0.53 & 0.37 & 0.30 & 0.23 \\ 
  3261654 & -45.603648 & 0.149636 & 17.01 & 15.81 & 15.83 & 15.87 & 15.93 & 3.68724 & 3.61116 & 0.97937 & 0.012 & 0.22 & 0.26 & 0.18 & 0.15 & 0.11 \\ 
  6058913 & -58.281094 & 1.205592 & 18.29 & 17.13 & 17.06 & 17.08 & 17.12 & 2.55444 & 2.55320 & 0.99951 & 0.040 & 0.48 & 0.50 & 0.35 & 0.30 & 0.26 \\ 
  6086465 & -57.509052 & -0.506643 & 18.44 & 17.26 & 17.31 & 17.38 & 17.43 & 3.63835 & 3.71848 & 0.97845 & 0.010 & 0.31 & 0.32 & 0.25 & 0.21 & 0.22 \\ 
  700313 & 15.188040 & -1.036745 & 17.85 & 16.72 & 16.56 & 16.50 & 16.51 & 2.99438 &  &  & 0.603 & 0.51 & 0.66 & 0.39 & 0.30 & 0.24 \\ 
  1731088 & -30.677761 & -0.062001 & 20.20 & 19.14 & 18.97 & 18.96 & 18.95 & 1.96133 &  &  & 0.357 & 0.87 & 0.91 & 0.66 & 0.51 & 0.49 \\ 
   \hline
\end{tabular}
\label{table:rrm}
}
\end{table*}

\vspace{6mm}
\begin{table*}
\caption{Multiperiodic RR Lyrae candidates with unusual period ratios in Stripe 82.  Notation is the same as for Table \ref{table:rrd}.}
{\scriptsize
\begin{tabular}{ccccccccccccccccc}
  \hline
ID & R.A. & Decl. & $u_{med}$ & $g_{med}$ & $r_{med}$ & $i_{med}$ & $z_{med}$ & $F_{\mathbf{z}_1}$ & $F_{\mathbf{r}}$ & Ratio & $P_{\mathbf{r}}$ & $A_{u}$ & $A_{g}$ & $A_{r}$ & $A_{i}$ & $A_{z}$ \\ 
  \hline
1528004 & -29.845729 & -0.438433 & 18.35 & 17.19 & 17.07 & 17.05 & 17.08 & 3.05227 & 5.03044 & 0.60676 & 0.020 & 0.87 & 0.92 & 0.67 & 0.53 & 0.46 \\ 
  3252839 & -46.822139 & -1.190310 & 17.83 & 16.69 & 16.70 & 16.75 & 16.80 & 3.21292 & 5.15024 & 0.62384 & 0.001 & 0.35 & 0.36 & 0.29 & 0.20 & 0.20 \\ 
  3218459 & -47.774708 & 0.387032 & 17.23 & 16.10 & 16.13 & 16.20 & 16.27 & 4.48794 & 3.60676 & 0.80366 & 0.000 & 0.15 & 0.17 & 0.12 & 0.10 & 0.08 \\ 
   \hline
\end{tabular}
\label{table:rro}
}
\end{table*}

The list of the 68 RRab and 36 RRc type objects is given in Table \ref{table:rrall}, containing the identifier, coordinates, median magnitudes corrected for interstellar extinction, the frequency and the bandwise amplitude estimates derived from a B-splines fit to the folded light curve. A summary plot for one of them is presented in Fig. \ref{fig:rrab}. For the ease of eventual further analysis and in order to present this addition to the sample of \citet{sesaretal10} in a coherent way, we fitted this new set of stars with the templates published there, and listed the template amplitude, the epoch of maximum brightness, the magnitude at the epoch of maximum brightness and the identifier of the best template in Table \ref{table:rrtempl}. There were several RRab stars that did not have acceptable template fits for all or part of the bands, as the template set is not all-encompassing  \citep{sesaretal10}. Such light curves showed a sharper rising branch, a narrower peak and a decreasing branch composed of a steep initial fading from the brightest state and a more prolonged near-minimum state than the templates. One example, the same star that is presented in Fig. \ref{fig:rrab}, is shown in Fig. \ref{fig:branifit}. The template amplitudes of these stars are underestimated on average by around 10\% (up to 20\%); such cases are denoted in Table \ref{table:rrtempl} by an asterisk following the amplitude value. For the RRab stars, we estimated heliocentric distance following Section 4.1.1 of \citet{sesaretal10}. We assumed the same average halo metallicity of [Fe/H] = -1.5 and a mean value $M_V = 0.6$ for their absolute magnitude. The extended heliocentric map of the halo is presented in Fig. \ref{fig:RRmap}. The majority of the new variables is situated close to the Galactic bulge, continuing smoothly the already known structures towards the bulge. 

The multifrequency analysis, performed on all candidates with sufficient residual degrees of freedom, yielded 23 further objects with multiple modes whose significance was confirmed by a combination of bootstrap and extreme-value procedures (though two with only very few residual degrees of freedom after smoothing), and two more showing clear light curve changes without indication of a second frequency.  These are listed  in Tables  \ref{table:rrd}, \ref{table:rrm} and  \ref{table:rro}. 

\begin{figure}
\begin{center}
\includegraphics[scale=.68]{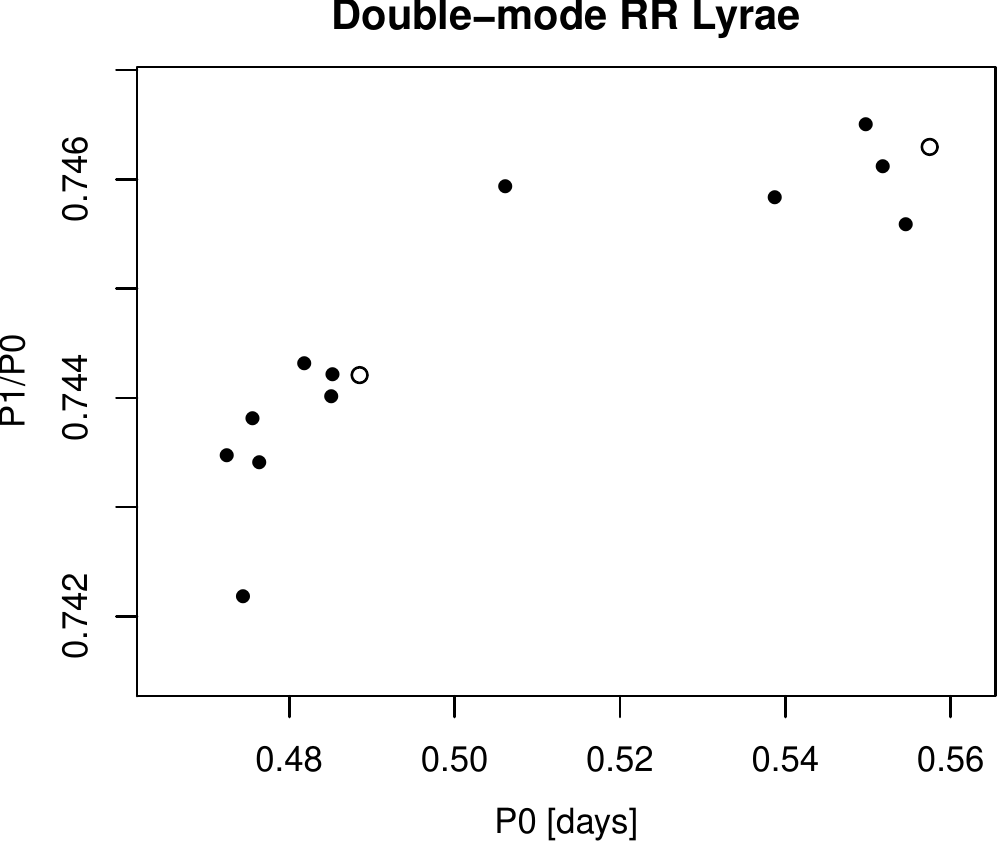}
\caption{Petersen diagram for the candidate double mode RR Lyrae variables. Black dots: candidates with well-identified frequency peaks, empty circles: aliased candidates, plotted here with the corrected periods.
}
\label{fig:p0-pratio}       
\end{center}
\end{figure}

Table \ref{table:rrd} contains the found double-mode RR Lyrae candidates showing the theoretically expected fundamental-overtone frequency ratio close to 0.745 (12 stars, around 2\% of the total Stripe 82 RR Lyrae sample found in \citealt{sesaretal10} and here). The summary information plot for one of them is presented in Fig. \ref{fig:doublemode}. These stars fit very well the expected relation between the fundamental period and the fundamental-overtone period ratio, as shown in Fig. \ref{fig:p0-pratio}. The span of periods indicates a broad range of metallicities and/or masses to account adequately for the observed ranges of periods and ratios (see e.g. \citealt{alcocketal00a}). In addition, Table \ref{table:rrd} lists two more objects for which the period search gave probably aliased results, and therefore the ratio is off for the frequencies corresponding to the maximum of the periodogram. However, supposing another alias to be the true frequency ($F_{\mathbf{r}} - 1$ day$^{-1}$ for the fundamental mode of object 1149344 and $F_{\mathbf{z}_1} + 1  $\; day$^{-1}$ for the overtone in the case of 1283137), the stars are in the regions admissible for double-mode RR Lyrae variables on the colour-colour, colour-period and period-amplitude diagrams. With these frequency choices, they fit well also the relationship between the fundamental period  and the period ratio, as shown by the empty circles in Fig. \ref{fig:p0-pratio}.

Table \ref{table:rrm} contains RR Lyrae candidates with two closely separated frequencies and/or showing multiple thread structure in their folded light curves, together with two others showing multiple threads without the presence of a resolved second frequency (an  example is given in Fig. \ref{fig:doublethread}). Such properties may be due to the Blazhko effect, slow periodic or irregular modulations of the amplitude or of the period. The folded light curve panels in Fig. \ref{fig:doublethread} indicate the Julian date of the observation with colours (the colour legend is given under the folded light curves), and the different threads visibly belong to different years of observations.  For this object, the separation of the two frequencies is very small, suggesting possibly a long Blazhko period, the presence of period changes or phase shifts. We list two more objects (6058913 and 6086465) with very small number of data where the smoothing left very few residual degrees of freedom, but the bootstrap indicated presence of secondary modes. More observations of these objects are needed before their type can be resolved. 

Several other RR Lyrae candidates  showed two significant peaks at neither closely spaced nor fundamen\-tal-first overtone  frequencies. Objects with similar frequency ratios were recently found by \citet{szosynskietal10} in OGLE data in the LMC and by \citet{olechmoskalik09} in $\omega$ Centauri. We present three of them in Table \ref{table:rro} for which the Monte Carlo assessment showed these secondary peaks significant. The first one, with its period ratio around 0.8, may be a candidate double-mode RR Lyrae pulsating in the first and second overtone (its summary information plots are given in the appendix, Fig. \ref{fig:RRf1-f2}); the other two show period ratios around 0.6. These objects need further data for confirmation.

\subsubsection{HADS candidates} \label{subsec_candsx}

\begin{table*}
\caption{ HADS candidates in Stripe 82.  Notation is the same as for Table \ref{table:rrall}.}
{\scriptsize
\begin{tabular}{ccccccccccccccc}
  \hline
ID & R.A. & Decl. &  $u_{med}$ & $g_{med}$ & $r_{med}$ & $i_{med}$ & $z_{med}$ & $F_{\mathbf{z}_1}$ & $A_{u}$ & $A_{g}$ & $A_{r}$ & $A_{i}$ & $A_{z}$ \\ 
  \hline
558620 & -12.512990 & -0.604203 & 20.50 & 19.42 & 19.46 & 19.50 & 19.56 & 23.22362 & 0.59 & 0.69 & 0.49 & 0.34 & 0.30 \\ 
  516958 & -13.867340 & 0.412679 & 17.37 & 16.40 & 16.32 & 16.37 & 16.43 & 17.68620 & 0.63 & 0.76 & 0.53 & 0.41 & 0.35 \\ 
  825896 & -16.982811 & -0.869013 & 17.57 & 16.41 & 16.33 & 16.37 & 16.43 & 20.17531 & 0.19 & 0.18 & 0.14 & 0.10 & 0.08 \\ 
  800977 & -18.396891 & 1.189213 & 18.21 & 17.21 & 17.17 & 17.20 & 17.26 & 19.90575 & 0.25 & 0.30 & 0.24 & 0.18 & 0.18 \\ 
  799691 & -19.685526 & -0.304110 & 20.58 & 19.63 & 19.61 & 19.65 & 19.67 & 19.11502 & 0.43 & 0.59 & 0.41 & 0.30 & 0.18 \\ 
  407172 & -8.778812 & -0.076337 & 20.63 & 19.67 & 19.64 & 19.71 & 19.79 & 26.77966 & 0.15 & 0.23 & 0.18 & 0.14 & 0.01 \\ 
  16959 & 0.818054 & 1.074244 & 20.52 & 19.41 & 19.45 & 19.51 & 19.49 & 16.82779 & 0.71 & 0.71 & 0.50 & 0.42 & 0.36 \\ 
  173268 & 10.194452 & -0.987410 & 20.01 & 19.02 & 18.86 & 18.83 & 18.85 & 16.12342 & 0.18 & 0.24 & 0.14 & 0.09 & 0.06 \\ 
  187850 & 13.118868 & -0.016244 & 20.79 & 19.78 & 19.70 & 19.75 & 19.79 & 19.19475 & 0.20 & 0.26 & 0.16 & 0.10 & 0.08 \\ 
  610306 & 17.691082 & 0.312217 & 20.60 & 19.59 & 19.57 & 19.63 & 19.69 & 19.46323 & 0.28 & 0.26 & 0.20 & 0.14 & 0.29 \\ 
  695598 & 18.277557 & -0.993741 & 21.00 & 19.92 & 19.91 & 19.97 & 20.07 & 20.49191 & 0.06 & 0.18 & 0.13 & 0.13 & 0.07 \\ 
  \vdots &  &  &  &  &  &  &  &  &  &  &  &  &   \\ 
   \hline
\end{tabular}
\label{table:sxall}
}
\end{table*}

\begin{table*}
{\scriptsize
\caption{Multiperiodic HADS stars in Stripe 82. Notation is the same as for Table \ref{table:rrd}.}
\begin{tabular}{ccccccccccccccccc}
  \hline
ID & R.A. & Decl. & $u_{med}$ & $g_{med}$ & $r_{med}$ & $i_{med}$ & $z_{med}$ & $F_{\mathbf{z}_1}$ & $F_{\mathbf{r}}$ & Ratio & $P_{\mathbf{r}}$ & $A_{u}$ & $A_{g}$ & $A_{r}$ & $A_{i}$ & $A_{z}$ \\ 
  \hline
4936224 & 58.031763 & 0.502842 & 17.71 & 16.81 & 16.81 & 16.85 & 16.94 & 17.91900 & 17.91776 & 0.99993 & 0.000 & 0.24 & 0.27 & 0.20 & 0.14 & 0.13 \\ 
  4433015 & -51.994003 & 1.211038 & 19.39 & 18.37 & 18.34 & 18.36 & 18.38 & 15.86637 & 16.06300 & 0.98776 & 0.008 & 0.30 & 0.26 & 0.17 & 0.11 & 0.13 \\ 
  2816955 & -44.450874 & 0.835961 & 16.95 & 15.95 & 15.97 & 16.06 & 16.14 & 25.80457 & 25.15836 & 0.97496 & 0.000 & 0.19 & 0.22 & 0.16 & 0.13 & 0.11 \\ 
  1113593 & -20.457095 & -0.775377 & 17.81 & 16.74 & 16.74 & 16.77 & 16.86 & 21.71685 & 21.04152 & 0.96890 & 0.000 & 0.23 & 0.29 & 0.21 & 0.17 & 0.14 \\ 
  1741727 & -31.076033 & -0.079779 & 20.06 & 19.01 & 18.95 & 18.99 & 19.05 & 20.68204 & 19.11928 & 0.92444 & 0.000 & 0.02 & 0.17 & 0.12 & 0.11 & 0.10 \\ 
  3642254 & -47.102871 & 0.553778 & 17.37 & 16.30 & 16.33 & 16.40 & 16.47 & 21.67150 & 18.67376 & 0.86167 & 0.001 & 0.23 & 0.30 & 0.20 & 0.16 & 0.15 \\ 
  5401947 & -55.894772 & 0.624779 & 16.67 & 15.60 & 15.62 & 15.67 & 15.78 & 11.90623 & 9.89844 & 0.83137 & 0.001 & 0.61 & 0.67 & 0.49 & 0.37 & 0.33 \\ 
  3269918 & -45.136352 & -1.230279 & 18.67 & 17.63 & 17.68 & 17.75 & 17.84 & 19.09286 & 24.45904 & 0.78061 & 0.000 & 0.46 & 0.49 & 0.37 & 0.28 & 0.24 \\ 
  2196466 & -35.874290 & -0.357666 & 19.32 & 18.30 & 18.19 & 18.19 & 18.23 & 9.31067 & 11.95100 & 0.77907 & 0.007 & 0.17 & 0.22 & 0.14 & 0.10 & 0.09 \\ 
  2777345 & -42.763135 & 0.482715 & 17.94 & 16.93 & 16.83 & 16.84 & 16.90 & 18.57330 & 24.30612 & 0.76414 & 0.003 & 0.21 & 0.24 & 0.18 & 0.12 & 0.09 \\ 
  2383752 & -37.431381 & -0.842875 & 21.33 & 20.31 & 20.27 & 20.30 & 20.32 & 13.15213 & 9.90024 & 0.75275 & 0.009 & 0.25 & 0.40 & 0.27 & 0.20 & 0.18 \\ 
  \vspace{3mm}
  5415273 & -57.074003 & -0.422064 & 18.34 & 17.30 & 17.25 & 17.30 & 17.35 & 14.23765 & 10.29948 & 0.72340 & 0.000 & 0.16 & 0.19 & 0.13 & 0.10 & 0.09 \\ 
  713584 & -16.915188 & 0.737187 & 19.40 & 18.37 & 18.38 & 18.41 & 18.48 & 20.72999 & 20.92128 & 0.99086 & 0.015 & 0.21 & 0.26 & 0.19 & 0.15 & 0.12 \\ 
  2298258 & -36.580453 & 0.002215 & 20.43 & 19.49 & 19.50 & 19.57 & 19.62 & 20.24135 & 19.23308 & 0.95019 & 0.044 & 0.39 & 0.49 & 0.36 & 0.29 & 0.24 \\ 
  421829 & -12.173045 & 0.426865 & 20.79 & 19.78 & 19.84 & 19.94 & 20.05 & 23.13966 & 25.14084 & 0.92040 & 0.034 & 0.16 & 0.22 & 0.17 & 0.11 & 0.07 \\ 
  225151 & 10.458869 & -0.877921 & 17.94 & 16.84 & 16.87 & 16.91 & 16.98 & 18.25562 & 22.35908 & 0.81647 & 0.015 & 0.45 & 0.52 & 0.37 & 0.30 & 0.25 \\ 
  4064319 & 5.074867 & -0.590467 & 17.95 & 16.90 & 16.90 & 16.98 & 17.08 & 21.84638 & 28.84696 & 0.75732 & 0.029 & 0.42 & 0.56 & 0.41 & 0.32 & 0.26 \\ 
  4377712 & -50.876368 & -1.088706 & 18.20 & 17.19 & 17.10 & 17.13 & 17.21 & 18.08047 & 24.00952 & 0.75305 & 0.040 & 0.60 & 0.70 & 0.50 & 0.40 & 0.34 \\ 
  3466895 & -46.531706 & -0.295154 & 20.75 & 19.75 & 19.73 & 19.81 & 19.92 & 21.52095 & 29.40172 & 0.73196 & 0.024 & 0.15 & 0.28 & 0.17 & 0.13 & 0.07 \\ 
  635626 & 15.536793 & -0.854048 & 21.04 & 20.14 & 20.01 & 19.99 & 19.99 & 18.50370 & 25.39660 & 0.72859 & 0.011 & 0.02 & 0.22 & 0.18 & 0.16 & 0.27 \\ 
  3482070 & -47.757087 & -1.185417 & 17.29 & 16.28 & 16.32 & 16.41 & 16.52 & 25.81594 & 16.93232 & 0.65589 & 0.029 & 0.14 & 0.16 & 0.16 & 0.09 & 0.06 \\ 
  2211584 & -35.050797 & 0.696938 & 20.82 & 19.89 & 19.84 & 19.89 & 19.95 & 17.68536 & 27.28028 & 0.64828 & 0.028 & 0.31 & 0.25 & 0.17 & 0.20 & 0.02 \\ 
  2950268 & -43.743677 & 0.280420 & 19.97 & 18.97 & 18.90 & 18.94 & 19.00 & 15.42760 & 3.00272 & 0.19463 & 0.030 & 0.41 & 0.44 & 0.29 & 0.20 & 0.21 \\ 
   \hline
\end{tabular}
\label{table:sxm}
}
\end{table*}

Using the training set described in Section \ref{subsec_presel}, the Random Forest procedure selected 163 candidates. This was reduced by a subsequent visual inspection to 132 accepted HADS candidates, 109 monoperiodic and 23 multiperiodic stars. These variables often have frequency spectra with complex structure, usually several secondary peaks beside a highly dominant frequency. The significance of the secondary peaks was much less prononunced in the multiperiodic HADS sample than among the double-mode RR Lyraes. Table \ref{table:sxall} presents a summary of the basic properties of the monoperiodic sample, similarly to Table \ref{table:rrall}. Two examples, one with symmetric and one with asymmetric folded light curve are presented in Figs. \ref{fig:sxsym} and \ref{fig:sx}. All frequencies given in Table \ref{table:sxall} can be affected by aliasing, though this does not modify the classification of the stars, as the typical frequencies of the HADS are high, and the aliasing would hardly lead to RR Lyrae frequencies and a consequent misclassification. In the absence of metallicity estimates or spectroscopic information, their further division into Population I ($\delta$ Scuti) and Population II (SX Phoenicis) objects is not possible.

In this sample, we also found several suspected multiperiodic objects. We included a star into our multiperiodic set only if two conditions held: (i) the combined bootstrap and extreme-value methods furnished a $P$-value smaller than 0.05 for the peak in the residual spectrum, (ii) the visual check of the folded residual light curve showed perceptible systematic variation rather than several outliers grouped by the folding. The properties of these stars are summarized in Table \ref{table:sxm}. We divided the table into two parts, the top half containing 12 objects with residual peak more significant than 0.01, and the bottom half another 11 with secondary peaks with $P$-value between 0.01 and 0.05. Among our multiperiodic candidates, we observed a broad variety of ratios between the primary and the residual periods at various levels of significance: they range from around 0.65 to 0.86, in addition to a group at very high ratios ($>0.92$), and one peculiar object with a ratio 0.19. The scatter of these values suggest a wide variety of objects of diverging types and with large differences in their characteristics such as the evolutionary state, mass, metal content or rotation. The majority of the most significant secondary peaks seem to form close doublets with the main frequency. Such behaviour is quite frequent among low-amplitude $\delta$ Scuti and SX Phoenicis stars (see e.g. \citealt{poretti03, bregerbischof02,mazuretal03}). However, aliasing makes it difficult to find the true frequency in both principal component and residual spectra, and the presence of a strong secondary peak may be also the consequence of a not sufficiently precise primary frequency identification due to the relatively scarce number of observations per time series. A few stars of our sample exhibits signs of amplitude or phase/period changes, light curve threads separated by observational year and showing a high dispersion of the peak values (the clearest example is 225151, shown in Fig. \ref{fig:doublemodeSX}).

\section{Discussion} \label{sec_disc}

Multi-filter observations comprise much information about the characteristics of the observed star through the colour variations. We tested robust PCA as a way to extract information about these variations. We found that PCA produces several useful quantities: the proportion of the variance of the first principal component to the total variance, the time series of the first principal component, a robust notion of outlyingness, and the direction of the first principal component.

We found that the proportion of PC1 variance to the total variance is a useful variability indicator. The PC1 spectrum can help detect cases when excess variance is due to underestimated errors, by having one dominant and four near-zero elements. The time series of the PC1 yields the best attainable signal-to-noise ratio. Also, weights for period search can be defined based on a robust notion of outlyingness.  Simulations and a test on real data containing several known RR Lyrae stars confirmed that period search on the PC1 time series with robust weights decreases the impact of aliasing on the results.

The direction of the largest variation in the 5-dimensional point cloud, which is termed the PC1 spectrum, proved to be useful in classification. For our sought sample of radially pulsating variables between 6500-7500 K, combined with bandwise SDSS error patterns, this direction points towards the $g$ band, with decreasing contribution from $r$, $i$, $u$ and $z$ bands. It was used in the Random Forest classification procedure as a new attribute helping to separate pulsating variables from eclipsing binaries. The coefficients of the $g$ and $i$ bands ranked among the best five attributes, beside the period and the $r-i$ and $g-r$ colours.

The study produced 132 HADS, 68 RRab, 36 RRc and 25 multiperiodic or peculiar RR Lyrae variable stars. The RR Lyrae are new addition to the confirmed RR Lyrae list of \citet{sesaretal10}. The vast majority of these new RR Lyrae stars were found in the regions not originally considered by \citet{sesaretal10}. Only a few type-ab RR Lyrae stars were missed by their work in the range considered by that study (308 deg $<$ RA $<$ 60 deg and $|$Dec$|$ $<$ 1.25), making their sample about 99\% complete. Among the new multiperiodic RR Lyrae, there are 14 double-mode candidates, several others showing signs of Blazhko effect, period or phase change, and we found candidates with unusual period ratios, of which one may be a double-mode star pulsating in the first and second overtone. Among the HADS candidates, the multiperiodicity seems to show a broad variety of ratios between periods. The HADS candidates, similarly to the multi-mode RR Lyrae, need eventual observational follow-up to clarify and confirm their type and pulsation modes.

This work yielded promising results about the utility of PCA, and opens the way to further improvements. Missing data reduce the number of observations in the time series of the first principal component to the number of data in the most scarcely observed band. Thus, though signal-to-noise ratio improves with the application of PCA, the decrease in the number of observations in the time series can imply worse performance of period search.  Optimization of the period search performance can be achieved either by restricting the analysis to the combination of only the best bands, or by a statistical methodology that is able to deal with the missing values. Further investigations will explore these interesting possibilities.

\section*{Acknowledgment}

M.S. wishes to thank J. Jur\-csik and G. Kov\'acs for helpful discussions. The work was supported by the Swiss National Science Foundation MHV grant  PMPDP2\_129178. 

Funding for the SDSS and SDSS-II has been provided by the Alfred P. Sloan Foundation, the Participating Institutions, the National Science Foundation, the U.S. Department of Energy, the National Aeronautics and Space Administration, the Japanese Monbukagakusho, the Max Planck Society, and the Higher Education Funding Council for England. The SDSS Web Site is http://www.sdss.org/.

The SDSS is managed by the Astrophysical Research Consortium for the Participating Institutions. The Participating Institutions are the American Museum of Natural History, Astrophysical Institute Potsdam, University of Basel, University of Cambridge, Case Western Reserve University, University of Chicago, Drexel University, Fermilab, the Institute for Advanced Study, the Japan Participation Group, Johns Hopkins University, the Joint Institute for Nuclear Astrophysics, the Kavli Institute for Particle Astrophysics and Cosmology, the Korean Scientist Group, the Chinese Academy of Sciences (LAMOST), Los Alamos National Laboratory, the Max-Planck-Institute for Astronomy (MPIA), the Max-Planck-Institute for Astrophysics (MPA), New Mexico State University, Ohio State University, University of Pittsburgh, University of Portsmouth, Princeton University, the United States Naval Observatory, and the University of Washington.

\appendix

\section[]{Summary information figures for visual checks}

These figures summarize the most important features of a star that can be derived from 5-band photometric time series. 

The two plots on the top left-hand side contain the frequency spectrum of the time series of $z_{11}, \ldots, z_{1T}$ and that of the residual PC1 time series, with the found most significant frequencies in each emphasized by an orange line. 

The three top panels in the middle column present the light curves of PC1 scores, $g$-magnitudes and $g-i$ colour  folded with the main frequency, and the two top right panels show the light curves of the PC1 and $g$ residuals folded by the residual frequency. Spline smoothed lines are added to the PC1 and $g$-band light curves. The folded light curves are colour-coded according to the Julian date of the observation; this is particularly useful for detecting slow phase or amplitude changes. The colour code is given in a stripe under the residual light curves. 

In the bottom row, we show the raw time series of observations, the $(u-g)$-$(g-r)$, the $(g-i)$-log$_{10}$(period) and the period-amplitude diagrams, and the PC1 spectrum. The raw time series in the leftmost bottom panel show the $u$-, $g$-, $r$-, $i$- and $z$-bands in blue, green, red, brown and black, respectively; time is given in Julian Dates. This panel is useful for detecting trends, shifts or  eventual other deviations in the data that might cause problems in the period search. The second bottom panel is the  $(u-g)$-$(g-r)$ colour-colour diagram. Light grey points here represent the general variable sample obtained by the condition imposed on the variance $d_1^2$ of the first principal component and the cuts on the PC1 spectrum $\mathbf{v}_1$ (steps 1 and 2 of the preliminary selection procedure in Section \ref{subsec_presel}). Black circles and dark grey dots refer to the RRab and the RRc sample of \citet{sesaretal10}, respectively. The orange blob represents the candidate star. In the  $(g-i)$-log$_{10}$(period) and the period-amplitude diagrams (third and fourth panels in the bottom row), the light grey points represent only the candidate HADS and RR Lyrae variables (without distinction) that were selected by Random Forest. Black circles, dark grey dots and the orange blob have the same meaning as in the colour-colour diagram. For the amplitude-period diagram, we used the percentile-based estimate of the $g$ light curve range as described  in Section \ref{subsec_presel}. The last plot, the PC1 spectrum show the $\mathbf{v}_1$ values of the object as an orange line versus the effective wavelengths (in \AA) of the filters. The two grey lines here give the lower and the upper boundary of the band occupied by the 483 confirmed RR Lyrae stars of \citet{sesaretal10}.

 \begin{figure*}
\begin{center}
\includegraphics[scale=.75]{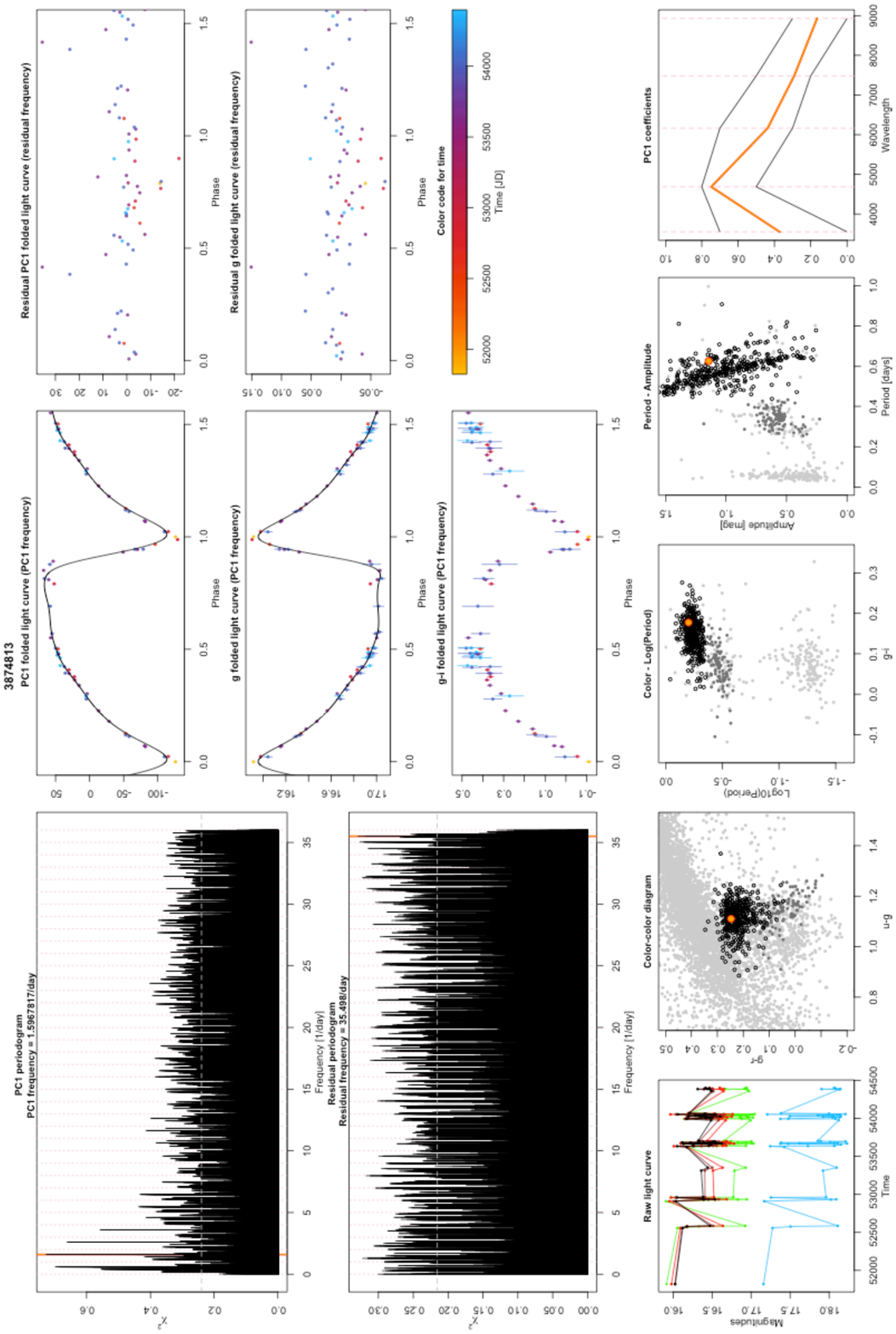}
\caption{An RRab candidate. Legend is given in the text of the Appendix.}
\label{fig:rrab}       
\end{center}
\end{figure*}

 \begin{figure*}
\begin{center}
\includegraphics[scale=.75]{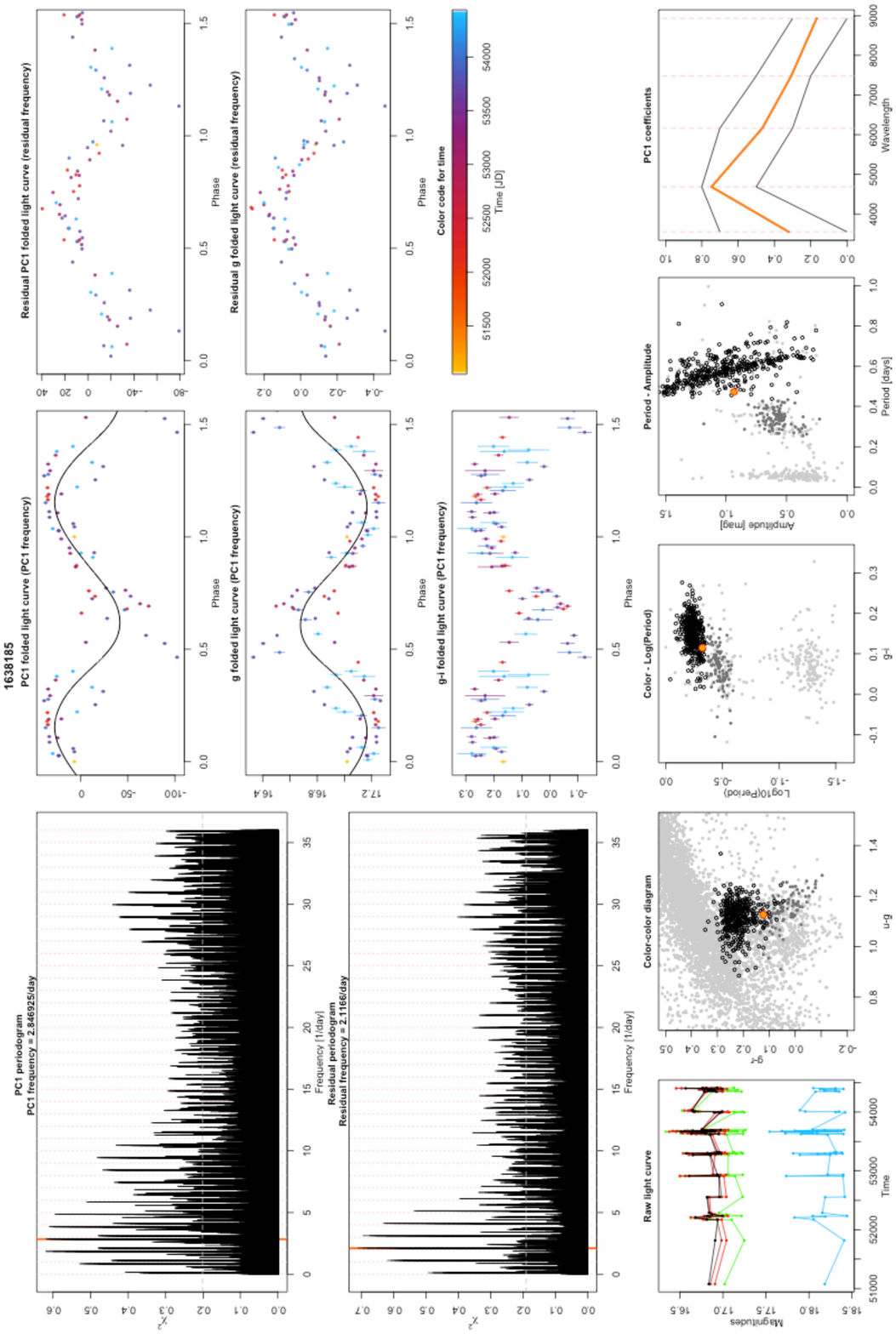}
\caption{A double-mode RR Lyrae candidate. Legend is given in the text of the Appendix.}
\label{fig:doublemode}       
\end{center}
\end{figure*}

 \begin{figure*}
\begin{center}
\includegraphics[scale=.75]{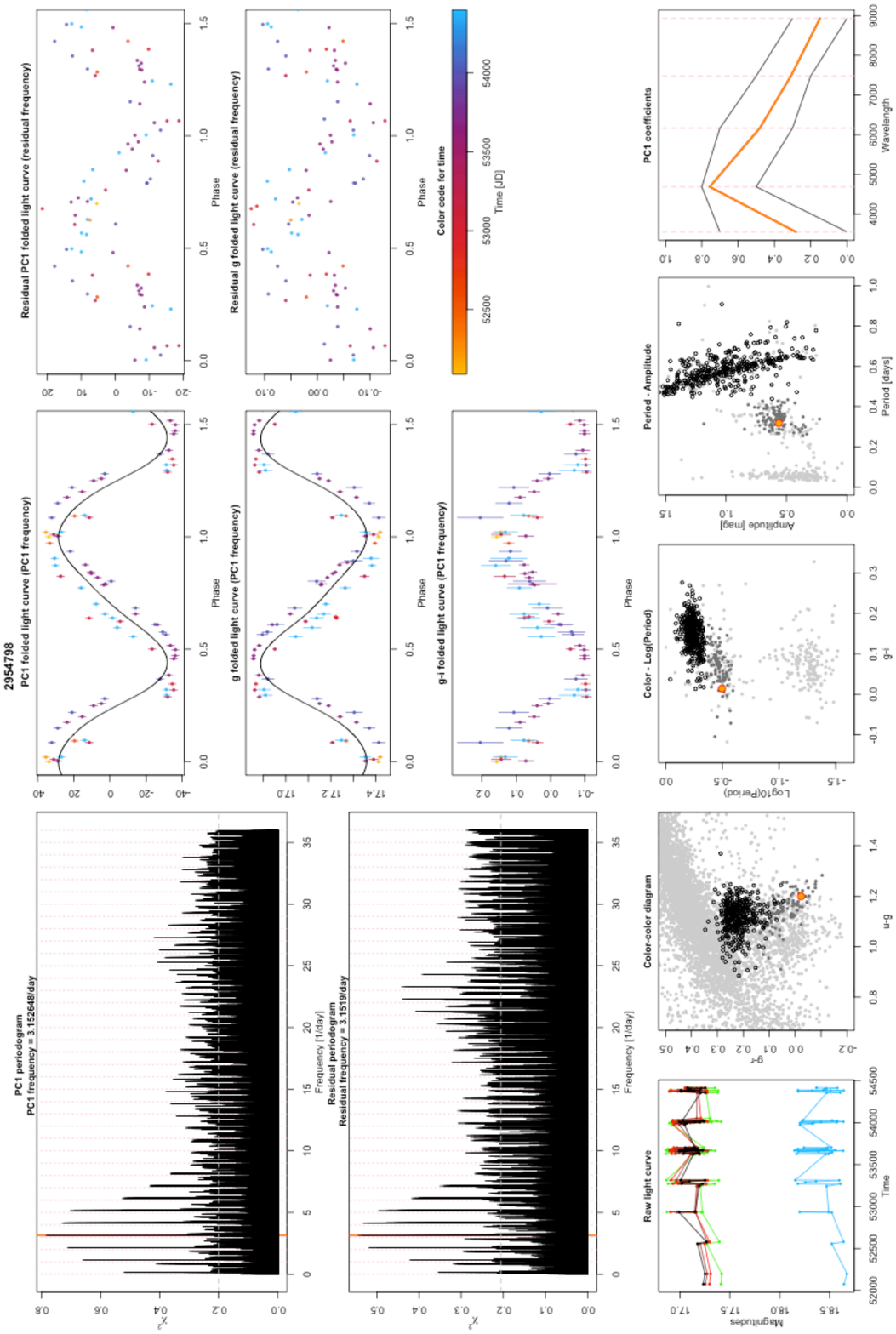}
\caption{An RR Lyrae candidate showing signs of phase and amplitude shift and possible Blazhko behaviour. Legend is given in the text of the Appendix.}
\label{fig:doublethread}       
\end{center}
\end{figure*}

 \begin{figure*}
\begin{center}
\includegraphics[scale=.75]{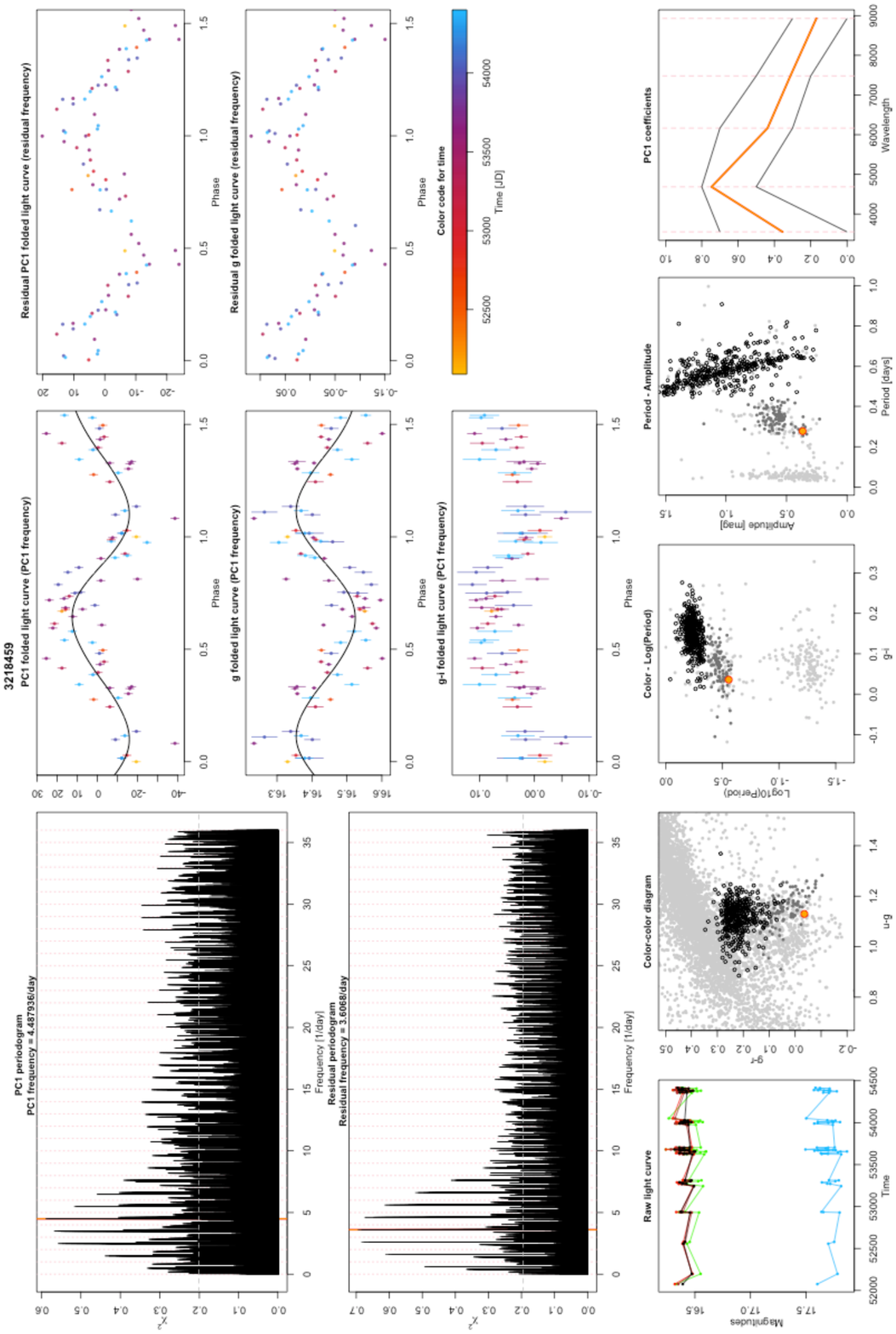}
\caption{A double-mode RR Lyrae candidate, possibly pulsating in the first and second overtone. Legend is given in the text of the Appendix.}
\label{fig:RRf1-f2}       
\end{center}
\end{figure*}

 \begin{figure*}
\begin{center}
\includegraphics[scale=.75]{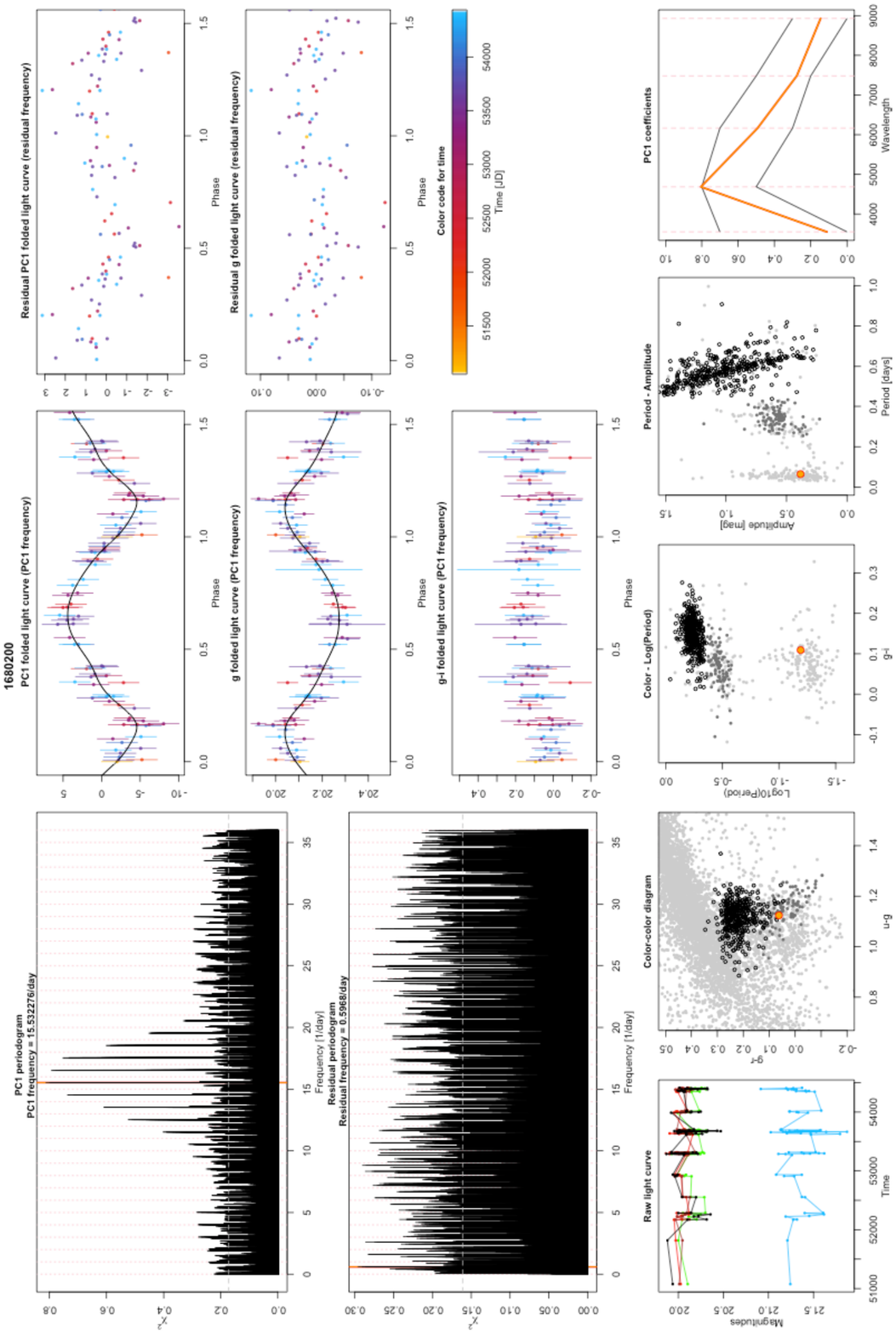}
\caption{A HADS candidate with symmetric light curve. Legend is given in the text of the Appendix.}
\label{fig:sxsym}       
\end{center}
\end{figure*}

 \begin{figure*}
\begin{center}
\includegraphics[scale=.75]{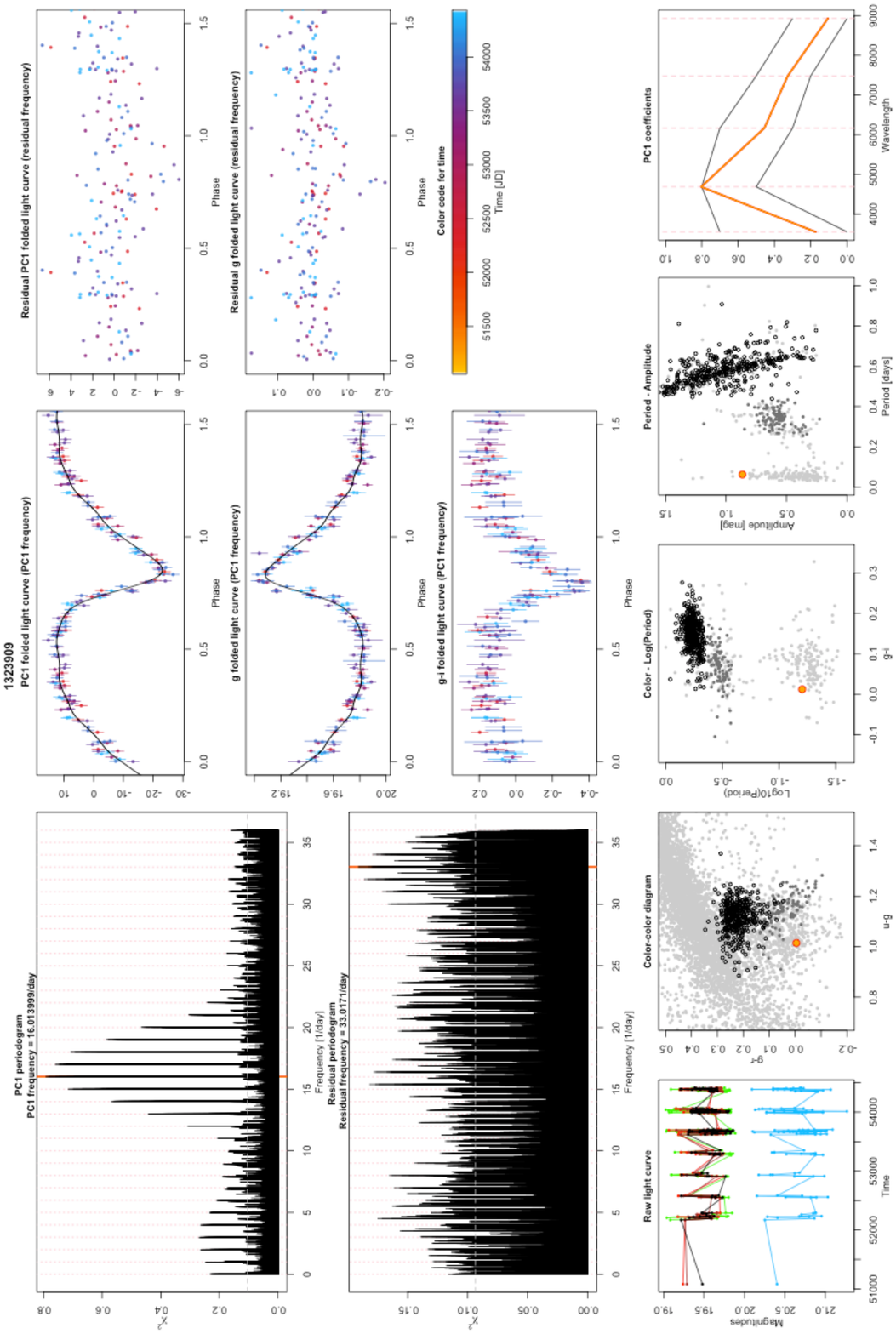}
\caption{A HADS candidate with asymmetric light curve. Legend is given in the text of the Appendix.}
\label{fig:sx}       
\end{center}
\end{figure*}

 \begin{figure*}
\begin{center}
\includegraphics[scale=.75]{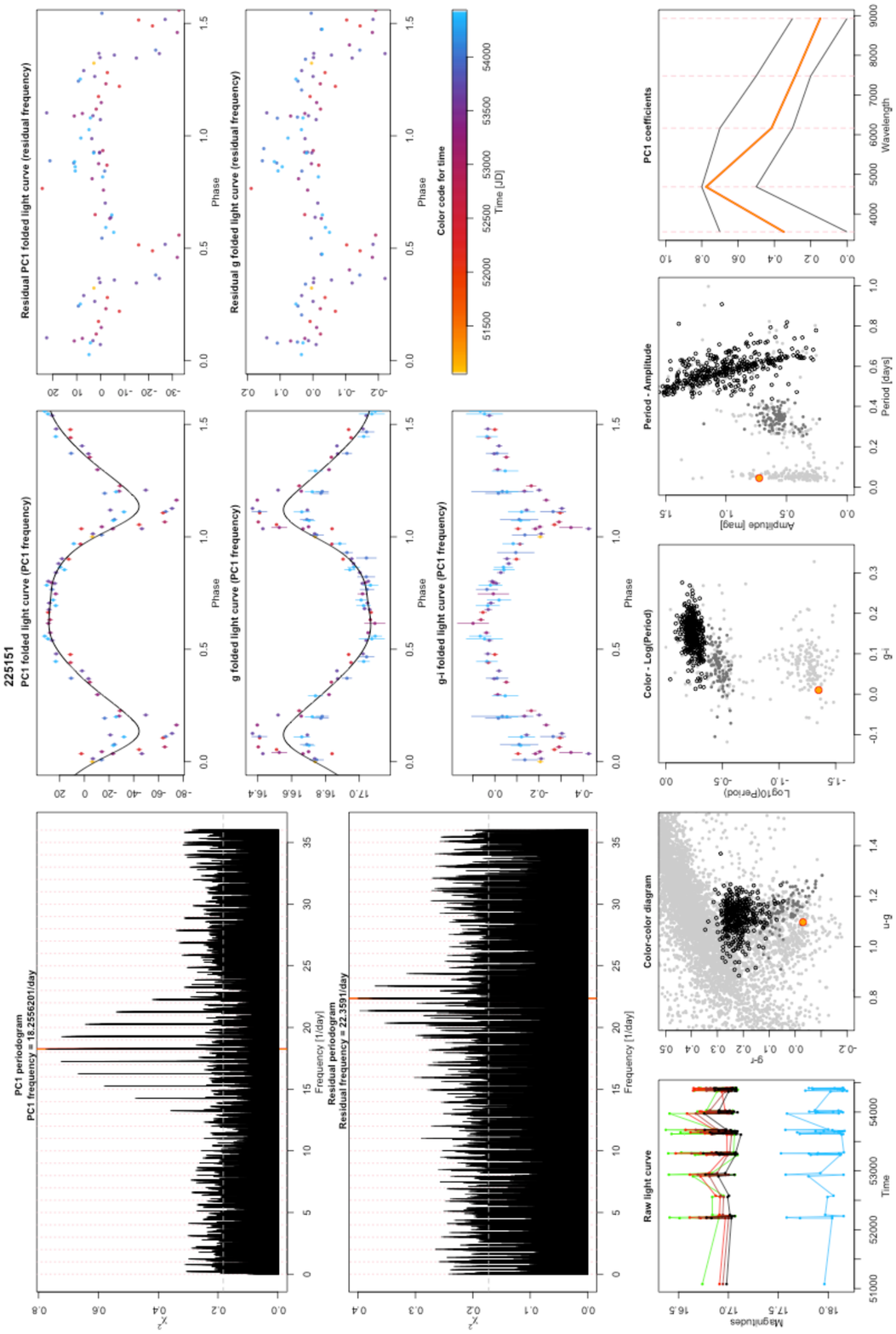}
\caption{A double-mode HADS candidate with amplitude change indications. Legend is given in the text of the Appendix.}
\label{fig:doublemodeSX}       
\end{center}
\end{figure*}

\label{lastpage}

\end{document}